\documentclass[global,twocolumn]{svjour}
\usepackage{graphics}
\usepackage{natbib}
\usepackage{xcolor}
\journalname{Rheologica Acta}
\begin{document}
\title{Microstructure and yielding of a capillary force induced gel}
\author{Sameer Huprikar \inst{1} \inst{3} \and Saurabh Usgaonkar
  \inst{1} \and
  Ashish K. Lele \inst{1} \and Ashish V. Orpe \inst{2} \inst{3}}
%
%
\institute{CSIR-National Chemical Laboratory, Pune 411008 India
\and Academy of Scientific and Innovative Research (AcSIR), Ghaziabad 201002 India}
\date{Received: date / Revised version: date}
%
\maketitle
\begin{abstract}
  We have investigated the rheology and structure of a gel formed from
  a mixture of non-Brownian particles and two immiscible liquids. The
  suspension of particles in a liquid undergoes gelation upon the
  addition of a small content of second, wetting liquid which forms
  liquid bridges between particles leading to a sample spanning
  network. The rheology of this gel primarily exhibits a yield stress
  at low shear rates followed by a linear variation of shear stress at
  high shear rates. The apparent yield stress extracted from the flow
  curves increases rapidly with volume fraction of second liquid
  before saturation, while it exhibits a monotonic increase with
  increasing particle concentration. Rescaling of the yield stress
  curves using suitable shift factors results in an empirical
  expression for the yield stress showing squared dependence on liquid
  fraction and a rapid increase with particle fraction above a certain
  value, both combined in a highly nonlinear manner. The
  microstructural variations with changing secondary liquid content
  and particle fractions are captured using three dimensional X-ray
  tomography technique. The microstructure is observed to show
  increased local compactness with increased liquid content and
  increased spatial homogeneity with increased particle fractions. The
  images from X-ray tomography are analysed to obtain the
  distributions of particle-particle bonds (coordination number) in
  the system which serve to explain the observed yield stress behavior
  in a qualitative manner.
\end{abstract}
\section{Introduction}
\label{intro}
The addition of a very small amount of an immiscible liquid to a
suspension of non Brownian hard sphere particles causes the suspension
to form a (solid-like) gel even at particle volume fractions lower
than $0.1$, which typically should exhibit liquid
behavior~\citep{kao75,cavalier02,mcculfor11,koos11,koos12a,koos12b,koos14a,koos14b,velankar14a,velankar14b,velankar15}. The
cause for this gel formation in most cases is attributed to the
wetting nature (interfacial contact angle $< 90$ degrees) of the added
secondary liquid with respect to the particle surface in the presence
of primary suspending liquid. The second liquid, thus, forms tiny
bridges between particles leading to a sample spanning network in the
system, called as a pendular state gel~\citep{koos11}. The solid-like
behavior arises due to attractive capillary force, comprising of
Laplace pressure inside the liquid and surface tension at the
liquid-liquid-solid contact line, between the particles which provides
resistance to flow for small enough forces and imparts a yield stress
to this gel.  Such solid like behavior is obtained even if the second
liquid is non-wetting (interfacial contact angle $> 90$ degrees). In
that case, particles cluster around the droplets of second liquid
causing a sample spanning network which is termed as capillary state
gel~\citep{koos11}.

Previous studies have shown these gels to exhibit a non-Newtonian
rheology comprising of yield stress ($\tau_{y}$) at low shear rates
($\dot{\gamma}$) and shear thinning behavior at high shear
rates~\citep{kao75,mcculfor11,koos12a,velankar15}.  The yield stress
represents solid-to-fluid transition~\citep{bonn17,coussot99} while
shear thinning behavior represents a decrease in viscosity ($\eta$)
with increasing shear. For gels arising primarily due to pendular
bridging between particles, the capillary attractive force between two
particles, separated by distance $s$, can be expressed by following
equation similar to that for wet, cohesive granular
materials~\citep{herming05,herming12}
\begin{equation}
  \label{eq:capillary-force}
  F_{c} = \frac{2 \pi R \Gamma \cos\theta}{1 + 1.05\bar{s} +
    2.5\bar{s}^{2}}
\end{equation}
where, $R$ is the particle radius, $\Gamma$ is the surface tension of
the added liquid, $\theta$ is the contact angle made by the liquid
with the particle surface, $\bar{s} = s \sqrt{R/V_{sl}}$ is the
normalised separation between particle surfaces and $V_{sl}$ is the
liquid bridge volume. The capillary force for particles with
negligible separation (i.e. particles touching each other) is given by
the well-known expression
$F_{c} = 2 \pi R \Gamma \cos\theta$~\citep{herming05}. The macroscopic
yield stress, incorporating the particle contact distribution and
number of particles per unit volume, is given
as~\citep{pietsch67,schubert84}
\begin{equation}
  \label{eq:yield_stress}
  \tau_{y} = f(\phi_{p}) g(\overline{V_{sl}}) \left(\frac{2 \pi
      \Gamma \cos\theta}{R} \right)
\end{equation}
where, $R$ is the particle radius, $f(\phi_{p})$ is a function of
particle volume fraction and $g(\overline{V_{sl}})$ is a function of
the normalised liquid bridge volume
($\overline{V_{sl}} = V_{sl}/R^{3}$) and can be expected to have a
constant value for a specified liquid fraction.  The yield stress
values obtained through stress ramp experiments by \citet{koos12a} for
varying temperatures scaled reasonably well with
eq.~(\ref{eq:yield_stress}), but by assuming
$f(\phi_{p}) \approx \phi_{p}^{2}$ as suggested previously for very
low volume fraction systems~\citep{pietsch67}.  Particles of different
types and in the diameter range of $1-20$ microns were employed in
this work. The value of $\Gamma$ was taken to be the interfacial
tension between two liquids and the two liquid contact angle
$\cos\theta$ calculated using the surface tension of individual
liquids.  A power-law dependence of yield stress on particle fraction,
but with an exponent of $3.3$, was also observed by \citet{velankar15}
for studies over a much wider range of volume fractions of silica
particles of $2$ micron diameter. The yield stress in their work was
obtained by fitting the well-known Herschel-Bulkley constitutive
equation to the steady shear rheology data. Using stress ramp
experiments, \citet{cavalier02} obtained a power-law dependence of
yield stress on particle fraction with an exponent of $4.6$ using
calcium carbonate particles of sub-micron size. The gelation mechanism
in their study was attributed to hydrogen bonding which was later on
shown to arise due to capillary forces~\citep{koos11}.

The magnitude of yield stress in several studies exhibits a rapid
increase with liquid content followed by a very slow
increase~\citep{cavalier02,koos11,koos12a,dittmann14,velankar15,bossler18}
and then a decrease at even higher secondary liquid
content~\citep{koos11,koos12a,velankar15}.  A similar dependence on
secondary liquid content, but for the shear viscosity, was observed
previously by \citet{mcculfor11} using glass spheres of $40$ micron
diameter. In a recent theoretical study~\citep{danov18,georgiev18},
the function $g(\overline{V_{sl}})$ in eq.~(\ref{eq:yield_stress}) was
expressed in terms of the balance between inter-particle capillary
attractive and electrostatic repulsive forces. The theory also
exhibited a power-law dependence of the yield stress on particle
fraction with an exponent of $0.66$ and was shown to agree with
experimental results for $4$ micron diameter particles.

Unlike the rheology studies which are quite substantial in number,
studies about the microstructural details of the gel systems are
relatively few and quite sporadic.  The formation of the most stable
structures for a capillary state gel was modeled by \citet{koos12b}
using optimisation codes for varying secondary liquid content and
contact angles. The increase and subsequent saturation of yield stress
with increase in liquid content was attributed, respectively, to the
presence of tetrahedral and octahedral clusters in the system. Using
confocal imaging, \citet{bossler16} investigated the effect of contact
angle (wettability) on the microstructure for solid as well as porous
particles. The transition from pendular bridging to a capillary state
gel was observed to shift to higher contact angles ($> 90$ degrees)
due to particle porosity. The microstructural differences between
capillary state and pendular state gel were quantified using the pair
distribution functions computed from acquired images.

In an another study, \citet{fortini12} used Brownian dynamics
simulations to obtain the evolution of the sample spanning network for
a capillary state gel starting from the formation of clusters to the
eventual percolation of these clusters. The fractal dimension
($d_{f}$) in their work was found to vary from $1.08\pm0.05$ for
initial cluster growth to a higher value of $2.6\pm0.1$ corresponding
to the random aggregation of very large sized clusters. The structural
changes for a pendular state gel, formed using particles of size $2$
micron diameter, were visualised using confocal microscopy by
\citet{velankar15}. The images showed the presence of percolated
pendular structure at lower liquid contents and that of large
aggregates at larger liquid contents. Using the observed power-law
behavior of yield stress on particle volume fraction and de Gennes
scaling concept for polymer gels, \citet{velankar15} obtained the
fractal dimension of their silica particle gel to be $1.79$, which is
close to that observed for diffusion limited
aggregation~\citep{weitz13}. In a very recent study,
~\citet{bossler18} obtained a range of values for fractal dimension
($1.9 - 2.8$) using rheological scaling laws and confocal imaging for
particles in the size range $0.5$ microns to $3$ microns. The varied
range of the values of fractal dimension were attributed to the
different lengthscales of the microstructure probed by individual
method, thereby, rendering non-uniqueness to the fractal dimension in
capillary gel systems. The fractal dimension, however, showed a
monotonic increase with particle size, suggestive of increasing
compactness in the gel system.

All the prior work presented above exhibits a reasonable body of work
available for capillary force induced gels, but centered around
particle sizes closer to the colloidal range. The primary intention of
this work is to extend the capillary gel literature to much larger
particle size ranges (few hundred microns), closer towards those
typically observed for the wet granular materials, which has not been
done previously. The primary reason for using larger particles is to
allow for much easier visualisation and quantification of the gel
microstructure, which has not been explored significantly in the
literature and which is absolutely necessary to understand the bulk
rheology behavior.  Secondly, this allows for the theoretical
advancement of the subject from the granular matter perspective as
well.  The larger sized particles would also preclude the Brownian
motion responsible to some extent for ageing, thereby simplifying the
rheology measurements~\citep{moller09,joshi18} and also reduce the
difficulties in measuring microstructural details due to possible
particle motion. While weaker gels are expected with increased
particle sizes~\citep{koos12a,bossler18}, the carryover of the entire
rheology behavior to gels formed for large sized particles is not
clear, which includes the dependence of yield stress on particle and
secondary liquid content. The rheology behavior of the system studied
over here reveals the presence of a yield stress at low shear rates
followed by a linear increase of shear stress at large shear rates.
Using suitable scaling analysis it is shown that the yield stress
exhibits a non-linear dependence on particle and secondary liquid
content.  The three dimensional visualisation of the microstructure
using laser imaging and X-ray tomography and subsequent quantification
of the microstructure using image analysis serves to explain the
qualitative nature of the yield stress behavior in a reasonable
manner.

\begin{table}
  \centering
  \caption{Physical properties of liquids used for gel
    preparation. CHB: Cyclohexyl bromide, DEC: Decahydronapthalene,
    TG: Thioglycerol and 1P: 1 propanol. The viscosities and surface
    tensions were measured, respectively, using rheometry and
    pendant drop method. The interfacial tension between the two
    liquid mixtures, CHB-DEC and TG-1P was measured as $3.36$
    mN$/$m. The static contact angle of TG-1P
    mixture on PMMA surface in the presence of CHB-DEC liquid was
    measured as $58 \pm 2.5$ degrees (see text for more details)}
  \label{liquid-properties}
  \begin{tabular}{ccccc}
    \hline\noalign{\smallskip}
    Liquid & Density & Viscosity & Refractive & Surface  \\
           & & & Index & tension \\
           & (g/cc) & (mPa s) & & (mN$/$m) \\
    \noalign{\smallskip}\hline\noalign{\smallskip}
    CHB & 1.33 & 2 & 1.495 & 29.9 \\
    DEC & 0.89 & 2 & 1.481 & 30.3 \\
    TG & 1.25 & 137 & 1.527 & 50.5 \\
    1P & 0.80 & 2 & 1.384 & 22.8 \\
    \noalign{\smallskip}\hline
  \end{tabular}
\end{table}

\section{Experimental details}
\label{exptl}
The suspension comprises of three constituents: (i) poly-methyl
methacryalate (PMMA) particles, purchased locally, of mean diameter
$D = 350 \pm 50$ microns as measured by imaging and analysis of
several hundred particles (ii) a primary suspending liquid made from a
mixture of two liquids, viz., Cyclohexyl bromide (CHB) procured from
Spectrochem and Decahydronapthalene (DEC) procured from Otto Chemicals
and (iii) secondary liquid also made from a mixture of two liquids,
viz., Thioglycerol (TG) procured from Sigma-Aldrich and 1-Propanol
(1P) procured from Thomas Baker. The physical properties for all the
four liquids are provided in Table~\ref{liquid-properties}.

The liquids, both, primary and secondary, were chosen such that (i)
the secondary liquid is wetting with respect to particle surface in
presence of primary liquid (ii) the densities and refractive indices
of the both liquids are same as that of the particles.  A perfect
density matching is needed to prevent the particles from
settling/floating in the suspending liquid (note the particles are
quite large in size while liquid viscosities are quite low). Particle
settling or floating leads to (i) undesired phase separation between
particles and liquids, which may hinder the preparation of the gel and
(ii) incorrect interpretation of the rheology data of the suspension
due to possible phase separation.  An accurate refractive index
matching is needed for visualisation in the bulk. It also helps in
ascertaining the accurate matching of density between particles and
liquid mixtures as explained later. The density of the particles and
liquids was matched to the order of $3-4$ decimal places, while the
refractive indices of particles and liquids was matched to make the
suspension transparent enough to visualize $20-30$ particle diameters
inside the sample. The details of refractive and density matching is
provided in~\ref{ri-den-match}.

\begin{figure}
  \centering \resizebox{0.45\textwidth}{!}{\includegraphics{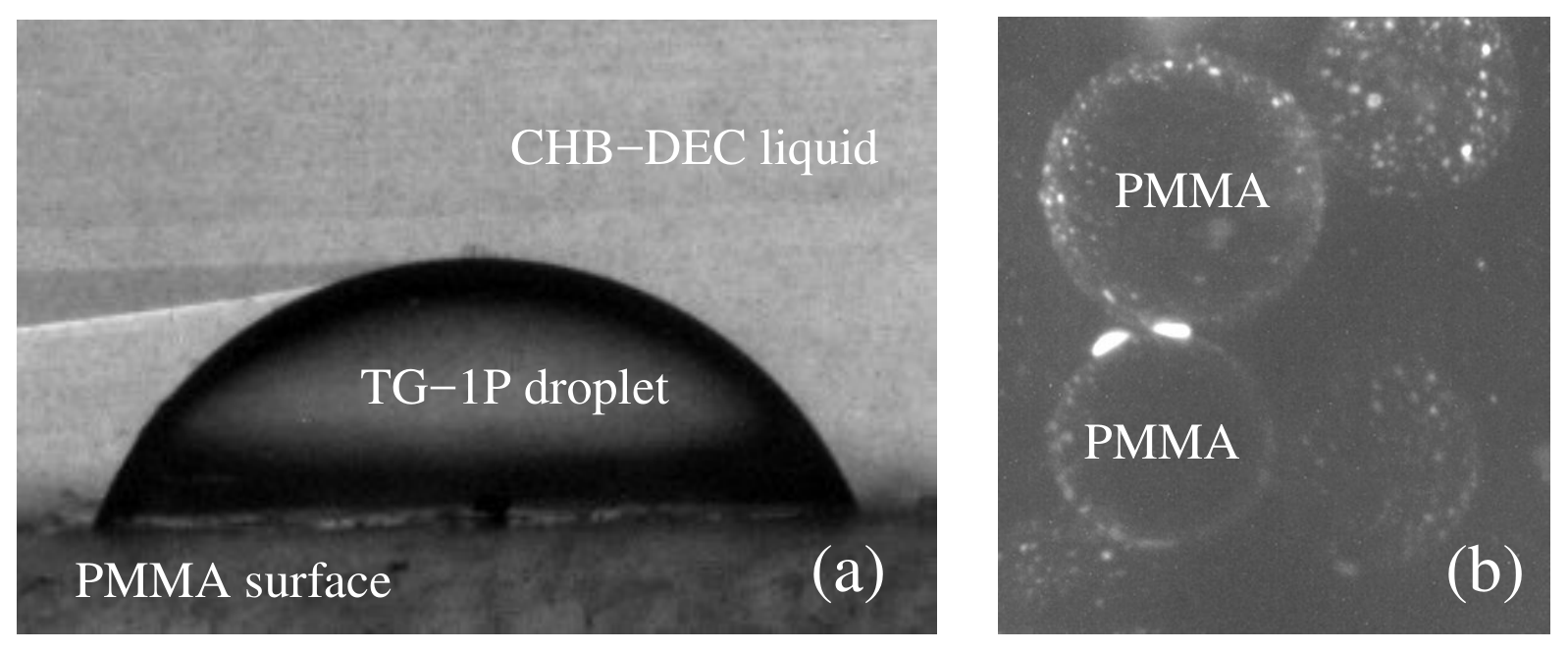}}
  \caption{(a) Contact angle formed by a drop of TG-1P mixture on the
    PMMA surface in presence of surrounding CHB-DEC mixture. (b)
    Formation of (TG-1P mixture) liquid bridge between two PMMA
    particles as well as droplet on particle surface in bulk CHB-DEC
    mixture. Two dimensional image about $5$ particle diameters away
    from the walls, obtained by fluorescing the secondary liquid using
    laser sheet. (See text for more details)}
  \label{contact-angle}
\end{figure}

The wetting ability of TG-IP mixture with respect to the particle
surface was measured in terms of the contact angle made by its drop on
a PMMA plate (of same material as particles) in presence of
surrounding CHB-DEC mixture. The drop was placed on a horizontal PMMA
plate immersed in CHB-DEC mixture and was imaged by a camera placed
sideways and in the plane of the plate (see
fig.~\ref{contact-angle}a). The images were analysed using ImageJ
software to obtain the static contact angle. The reported value of the
contact angle in Table~\ref{liquid-properties} represents an average
over ten independent measurements. Given the value of the measured
contact angle, it is expected that the TG-IP mixture, when added as a
secondary liquid to the suspension of PMMA particles in CHB-DEC
mixture, will form a gel arising out of pendular bridging between
particles (as clearly evidenced in fig.~\ref{contact-angle}b and
fig.~\ref{liquid-bridge-distribution}). Corresponding description is
provided in section~\ref{micro}).

Stress ramp measurements were carried out for every gel sample using
concentric cylinder (cup-bob) assembly in the stress controlled
MCR-301 rheometer (Anton Paar Inc.). Given the large particle size
employed, the outer cylinder of larger radius ($11.33$ mm) was
fabricated separately to employ shearing thickness ($3$ mm) of about
$8 -9$ particle diameters given the radius of the rotating bob to be
$8.33$ mm. In all the experiments the temperature within the rheometer
was maintained at $25 \pm 1$ deg C.  Instead of preparing the sample
and then transferring to the rheometer, the sample was prepared in the
detachable cylinder (cup) itself. Initially, a pre-determined quantity
of particles was added to the primary liquid (CHB-DEC mixture) and
stirred for about $10$ minutes. A known, tiny quantity of secondary
liquid (TG-1P mixture) was then added to the suspension and the entire
mixture was stirred again for another $10$ minutes. This second
stirring yielded a gel-like material whose strength and
characteristics are expected to depend on the quantity of particles
and secondary liquid added.  It is anticipated that the stirring of
the mixture generates several microdroplets from the secondary liquid
which leads to formation of the gel. Particle volume fraction
($\phi_{p}$), was varied from $0.1$ to $0.4$, while secondary liquid
volume fraction ($\phi_{sl}$) was varied from $0.001$ to $0.025$. The
volume fraction of an entity (particle or liquid) was obtained as the
ratio of its volume to the total volume comprising of both the liquids
and particles. The cylinder was then assembled back in the rheometer.
The gel was first rapidly sheared at $100$ $1/s$ for $2$ minutes to
break down the gel completely to a sol state having a very low
viscosity. The sample was then allowed to recover under quiescent
conditions for $20$ minutes (minimal stress applied for five minutes
followed by no stress for fifteen minutes) till the viscosity
increases significantly and reaches a constant value. This attainment
of the constant viscosity value represents the initial state of the
gel and the methodology, thus implemented, allows for complete removal
of any preparation history. The gel was then subjected to stress ramp
measurements, wherein the stress was increased in stepwise manner over
the entire range. The shear rate was recorded at each stress after
$10$ seconds of acquisition time ($t_{aq}$). A few experiments were
also done for $t_{aq} = 100$ seconds and $t_{aq} = 1000$ seconds. In
essence, each step is a creep experiment over the acquisition
time. The magnitude of applied stress was varied over four orders of
magnitude with the corresponding measured shear rate varying over
eight orders of magnitude. It should be noted that the formation of
the droplets of secondary liquid, hence the gel, is sensitive to the
preparation protocol which includes the constituent addition sequence
as well as the stirring speed of the mixture. The change in the
protocol, thus, may alter the observed rheology and the underlying
microstructure as shown previously~\citep{velankar14b,yang17}.

The three dimensional microstructural details of the gel were obtained
using X-ray tomography (Zeiss XRadia 510 Versa, X-ray microscope). A
small amount (about $1.5$ ml) of gel sample (obtained after stirring
as mentioned above) was transferred to a polypropylene microcentrifuge
tube (volume $2.5$ ml), which was carefully placed on the sample
holder to ensure centralised positioning. The assembly was placed in
the path of a polychromatic X ray beam and rotated by $360$ degrees.
An image of the gel sample (approximately $20 \times 20$ particle
diameters) was acquired every $1.2$ degrees of rotation at an exposure
time of $10$ seconds per image. The three dimensional reconstruction
of all images (region spanning $20 \times 20 \times 20$ particle
diameters) was then analysed using standard IDL routines to obtain the
centroid and radius of every particle (procedure described in
section~\ref{micro}) which were further used to determine relevant
structural information in the system.

\section{Results and Discussion}
\label{results}
The behavior of the flow curves (stress vs shear rate) obtained from
step-wise stress ramp rheology is discussed first. The value of the
yield stress for different cases is obtained from the flow curves and
its dependence on variation of $\phi_{p}$ and $\phi_{sl}$ is discussed
next.  This is followed by simple scaling analysis (data shifting) to
obtain an empirical expression for the yield stress in terms of
$\phi_{p}$ and $\phi_{sl}$. We, then, discuss the microstructural
details of the gel as acquired using three dimensional X-ray
tomography technique and provide some quantitative measurements to try
and explain the yield stress behavior.

\subsection{Stress ramp rheology}
\label{rheol}
\begin{figure}
  \centering \resizebox{0.4\textwidth}{!}{\includegraphics{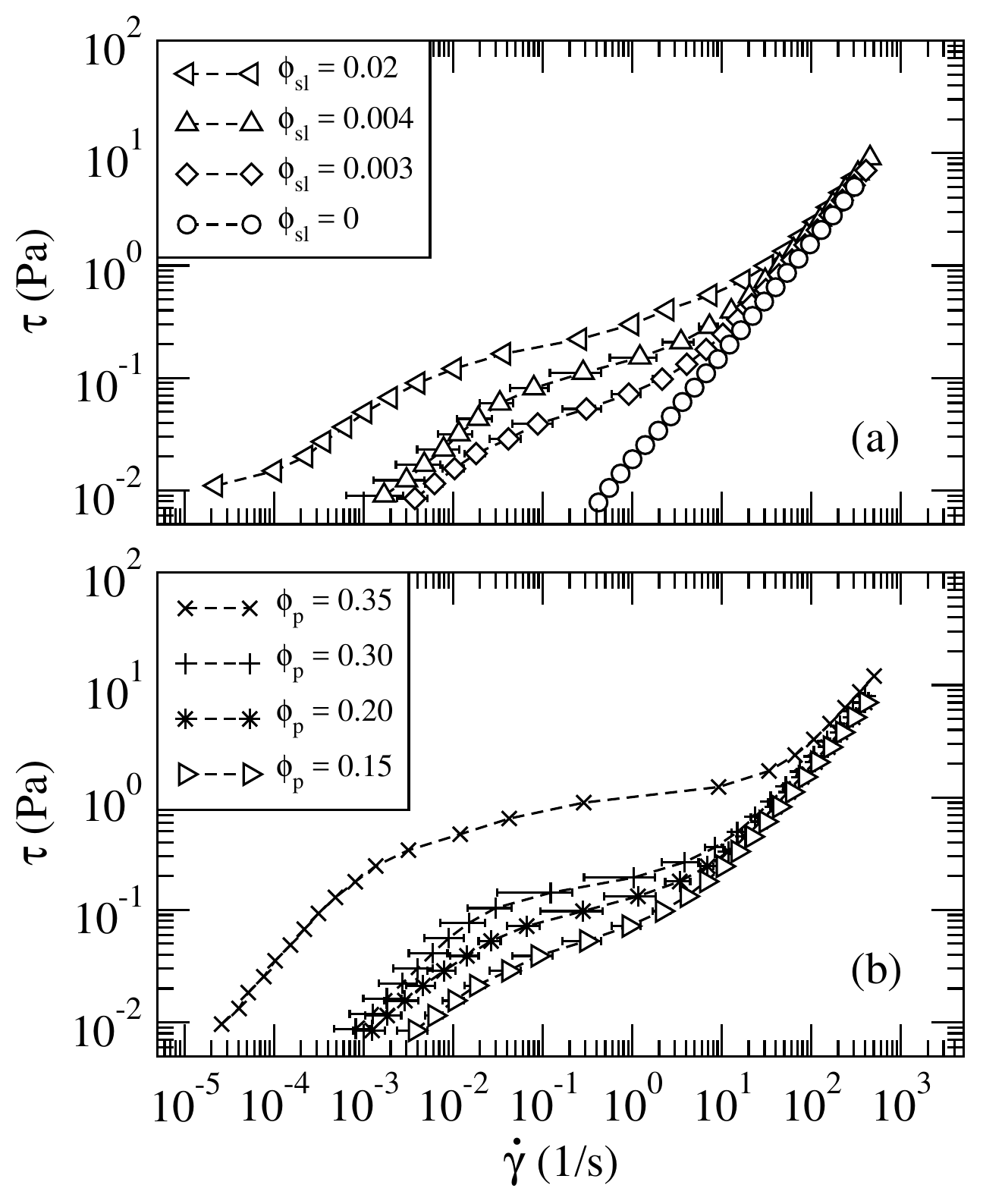}}
  \caption{Stress ramp rheology data. (a) Experimental data for
    $\phi_{p}=0.15$ (b) Experimental data for $\phi_{sl}=0.003$. Every
    curve shows three distinct flow regimes (see text for more details)}
  \label{steady-rheology}
\end{figure}

Figure~\ref{steady-rheology} shows the variation of shear stress
($\tau$) with shear rate ($\dot{\gamma}$) for varying values of
$\phi_{sl}$ and $\phi_{p}$. The error bars, shown for a few profiles
(corresponding to the images shown in fig.~\ref{gel-microstructure}),
represent the standard deviation over $6-7$ independent
measurements. Without the addition of secondary liquid
($\phi_{sl} = 0$), a linear behavior is observed throughout as would
be expected for the shearing of a dilute suspension of hard
spheres~\citep{powell05}. The addition of tiny amounts of secondary
liquid induces non-linear behavior (see
fig.~\ref{steady-rheology}a,b). A continuous increase in the shear
stress value is observed over $7-8$ orders of magnitude variation in
shear rates.

When subjected to very high shear rates ($>> 10 s^{-1}$), the gel
structure is expected to break, thereby reducing it predominantly to a
binary system of particle and primary suspending liquid, with
interspersed secondary liquid droplets. The flow curve should then
exhibit a behavior similar to that of a binary particle-primary liquid
suspension. All the flow curves for constant $\phi_{p}$ and varying
$\phi_{sl}$ (see fig.~\ref{steady-rheology}a), merge onto the curve
for $\phi_{sl} = 0$ at higher shear rates, thus indicating a linear
behavior for all cases with nearly similar viscosity values
($\eta_{\infty}$).  The flow curves for constant $\phi_{sl}$ and
varying $\phi_{p}$ (see fig.~\ref{steady-rheology}b) also merge onto
one curve at higher shear rates, again suggesting linear behavior and
constant viscosity. However, the viscosity values are slightly
different in each case, corresponding to the values of $\phi_{p}$, but
are not clearly discernible due to closeness of the data on a log-log
plot.  The viscosity values measured separately for a binary
suspension (no secondary liquid) show a very slow increase with
increasing $\phi_{p}$ till $0.4$ beyond which they increase rapidly.

For the lower shear rate range ($10^{-2} s^{-1}$ to $10 s^{-1}$), the
shear stress shows a very slow variation with shear rate (plateau or
shoulder formation in the flow curves). This represents the yield
stress part of the curve. Previous studies for very small particle
sizes have observed flow curves exhibiting two regions, viz. a true
yield stress followed by shear thinning behavior (Herschel-Bulkley
model) using polymeric primary as well as secondary
liquids~\citep{velankar15} and an ideal Bingham flow behavior using
Newtonian liquids~\citep{kao75}.

\begin{figure}
  \centering \resizebox{0.4\textwidth}{!}{\includegraphics{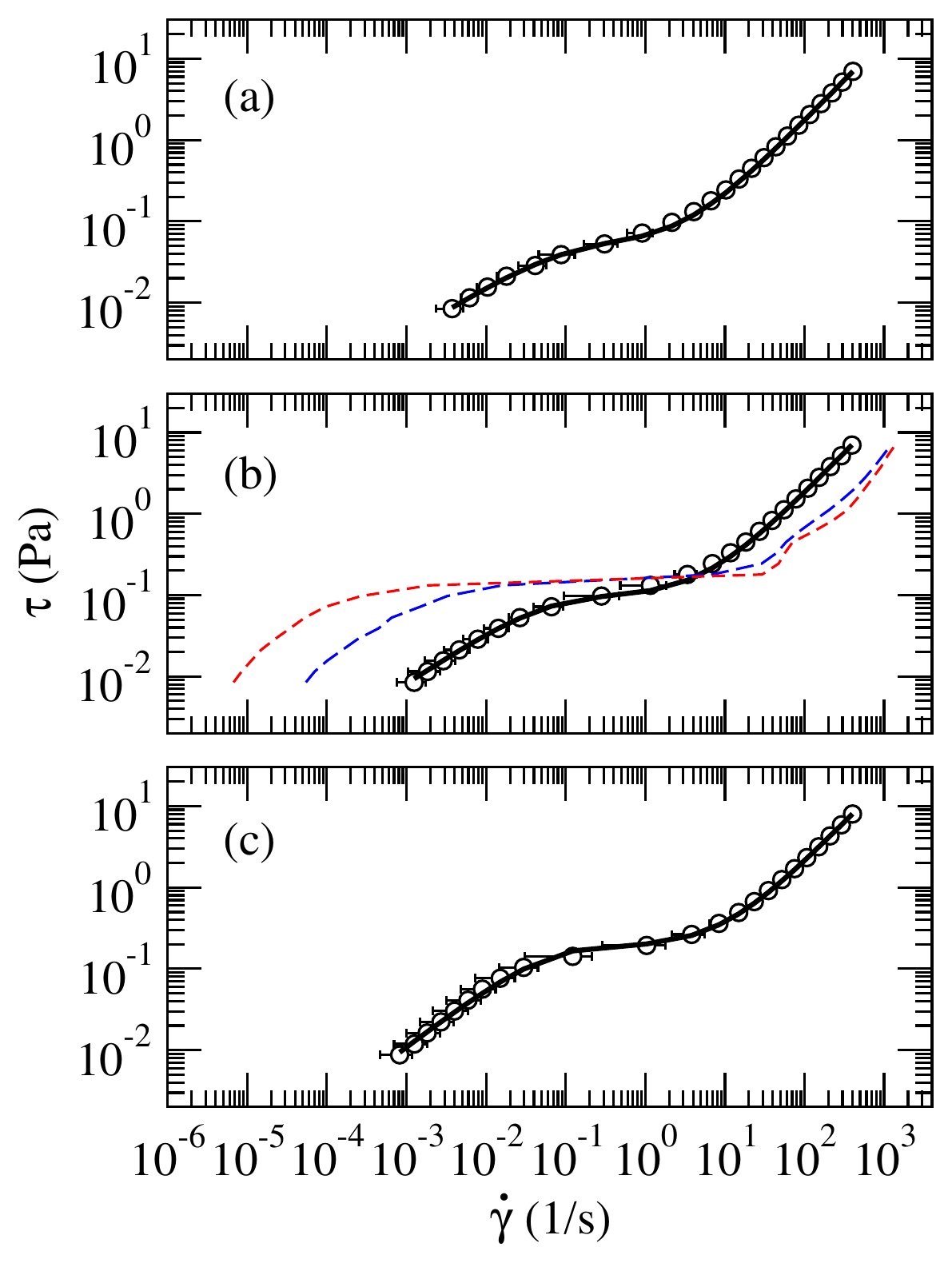}}
  \caption{Fitting of the Bingham-Papanastasiou model (see text for
    details) with the stress ramp rheology data obtained for
    $\phi_{sl} = 0.003$ and (a) $\phi_{p} = 0.15$ (b)
    $\phi_{p} = 0.20$ (c) $\phi_{p} = 0.30$. The dashed blue and red
    lines in (b) represent data for longer acquisition times
    ($t_{aq}$) of $100$ s and $1000$ s, respectively. (See text for
    more details)}
  \label{model-fit}
\end{figure}

The absence of true yield stress type behavior observed in
fig.~\ref{steady-rheology} is most likely due to the smaller
acquisition time ($t_{aq} = 10$ seconds) used in the stress ramp
experiments. For $t_{aq} = 100$ and $t_{aq} = 1000$ seconds, a more
pronounced plateau is observed. The flow curves below this plateau
shift to much lower shear rates than those observed for $t_{aq} = 10$
seconds. This behavior is shown for one case, $\phi_{p}: 0.20$ and
$\phi_{sl}: 0.003$, in fig.~\ref{model-fit}b. Similar qualitative
behavior is observed for all other cases studied. A rightward shift,
towards higher shear rates, is also observed, though the effect
saturates with increasing values of $t_{aq}$. The magnitude of the
yield stress, however, does not change with the change in acquisition
time. The observed behavior suggests that the non-Brownian suspensions
studied in this work are likely to exhibit true yield stress and
discontinuous flow curve, but only for large enough acquisition times.

The observed increase of shear stress with shear rates, below the
yield stress, can be attributed to subtle rearrangements within the
gel microstructure, which may be transitory in nature given the
dependence on acquisition time. It is, thus, not surprising that flow
curves, below the yield stress, are different from each other for
varying combinations of particle and secondary liquid content used.

With a view to further explore the nature of the gel, particularly
below yielding, the strain experienced by the sample was measured for
each stress value over a particular acquisition time used. The
variation of the applied stress with the realised strain and pertinent
description is provided in~\ref{stress-strain}. The reasonable
superposition of the data for different acquisition times below the
yielding is suggestive of a possible elastic deformation of the gel
which is conspicuously absent for stresses above yielding. The strain
itself, however, increases with time for stresses above as well as
below yielding (see the description and corresponding figure
in~\ref{strain-evol}). The increase in strain represents plastic
deformation of the sample in consideration. The complicated rheology,
so observed, for the gel below the yielding will require further
investigation which is not within the scope of present work.

To elucidate the possible effect of wall slip and confinement on the
rheology results, a few measurements were carried out in the
separately fabricated outer cylinder of larger radius ($14.33$ mm)
resulting in a larger shearing thickness ($6$ mm), about $18 - 20$
particle diameters wide. The rheology data and the corresponding
discussion for two shearing gaps is provided in~\ref{gap-depen}. The
results suggest that the overall rheology behavior does not seem to be
qualitatively affected by the shear gap and due to the presence of a
small amount of possible wall slip. Our primary interest is to
understand the yield stress behavior and its dependence on the
microstructure (presented in section~\ref{micro}).  The remainder of
the discussion refers to the data acquired in $3$ mm shearing gap and
for an acquisition time ($t_{aq}$) of $10$ seconds.

Given the absence of easily discernible yield stress in the flow
curves shown in fig.~\ref{steady-rheology}, it is not possible to fit
a Bingham constitutive equation to obtain the yield stress value. We,
instead, make use of the following phenomenological stress
constitutive equation proposed by ~\citet{papanastasiou87}, which has
been used significantly for modeling studies of polymeric
systems~\citep{mitsoulis07}
\begin{equation}
  \label{eq:papanas}
  \tau = \tau_{y} [1 - exp(-A\dot{\gamma}^{q})] + \eta_{\infty} \dot{\gamma}
\end{equation}
where, $\tau_{y}$ represents the yield stress, $\eta_{\infty}$
represents the Newtonian viscosity and $\dot{\gamma}$ represents the
shear rate. For $A = 0$, the equation for Newtonian liquid is
recovered, while for $A \rightarrow \infty$, the model is same as the
ideal Bingham visco-plastic model. The parameters $A$ and $q$ are
adjusted to obtain the best fit to the data. We do not observe any
specific trend in the values of $A$ and $q$, the values for which are,
thus, not reported over here.  We wish to stress here that the use of
eq.~(\ref{eq:papanas}) is done purely for the sake of mathematical
fitting in order to extract a value of yield stress from the flow
curve data. By using eq.~(\ref{eq:papanas}) we do not mean to imply
that the flow of the suspensions can be physical modeled by the
Papanastasiou constitutive equation. Indeed, the time dependent stress
response in the very low shear region is not captured by
eq.~(\ref{eq:papanas}). A more complex constitutive relation will be
required to capture the elasto-viscoplastic character of the
suspension described in \ref{stress-strain} and \ref{strain-evol}
which is not within the scope of the present work.  The fit of
eq.~(\ref{eq:papanas}) to the data is shown for a few cases in
fig.~\ref{model-fit}, which is representative of fits obtained for all
the data.  The yield stress ($\tau_{y}$) obtained from data fitting
shows interesting and rich behavior which will be discussed next.
 
\subsection{Yield stress behavior}
\label{yield}
\begin{figure}
  \centering \resizebox{0.4\textwidth}{!}{\includegraphics{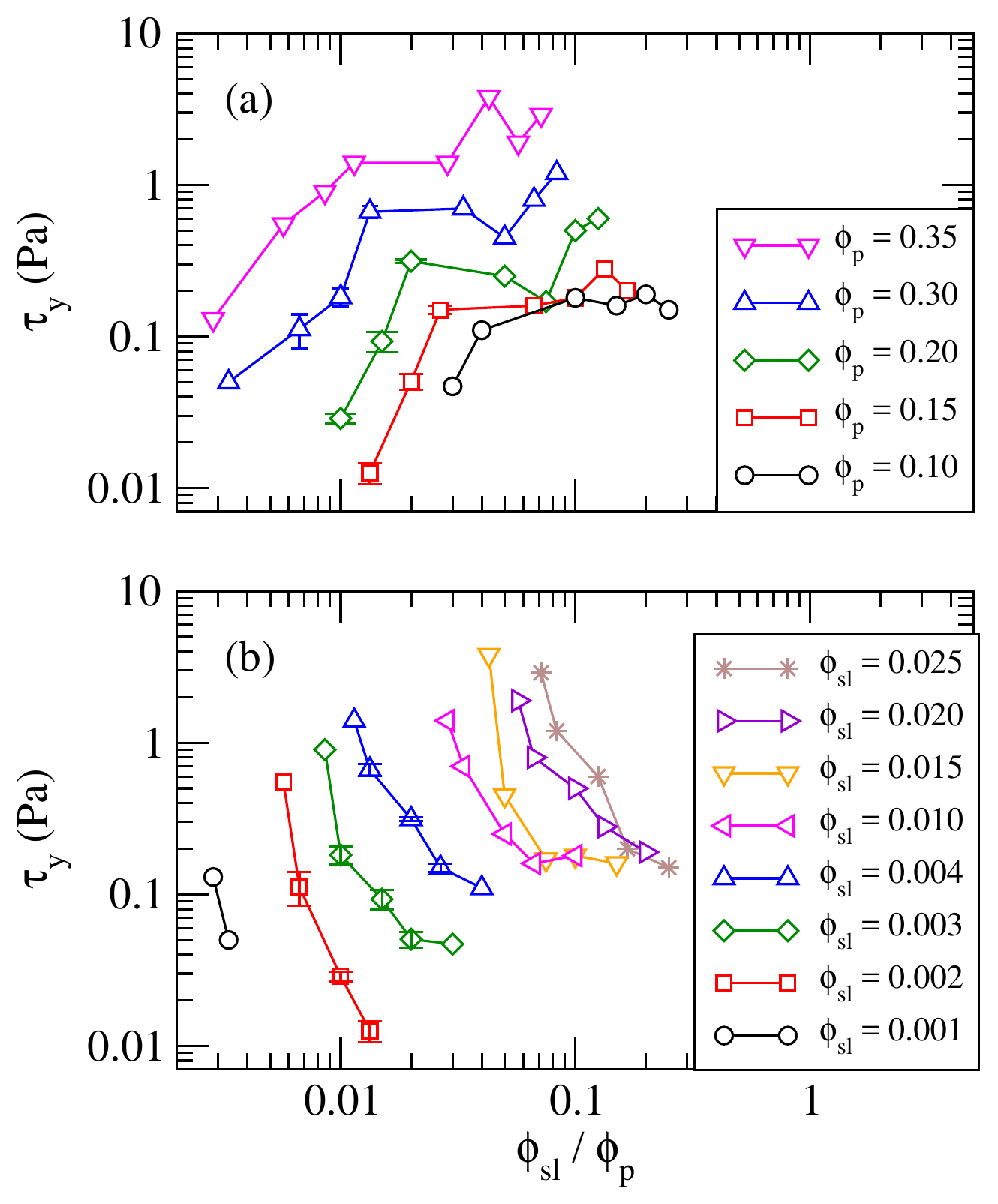}}
  \caption{Variation of yield stress ($\tau_{y}$) with the ratio
    $\phi_{sl}/\phi_{p}$. (a) Data for constant values of $\phi_{p}$
    and varying values of $\phi_{sl}$. (b) Data for constant values of
    $\phi_{sl}$ and varying values of $\phi_{p}$.}
  \label{yield-stress-phi-ratio}
\end{figure}

The variation of yield stress with the ratio $\phi_{sl}/\phi_{p}$ is
shown in fig.~\ref{yield-stress-phi-ratio} for the entire range of
$\phi_{sl}$ and $\phi_{p}$ investigated in the experiments. The error
bars shown for a few cases (corresponding to the images shown in
fig.~\ref{gel-microstructure}) represent the standard deviation of the
yield stress values obtained by fitting the model
(eq.~(\ref{eq:papanas})) to $6-7$ independent stress ramp flow
curves. They also correspond to those cases for which the
microstructure was studied in detail (described in
section~\ref{micro}). Four different behaviours are evident from the
figure.  The yield stress of the gel first increases rapidly with
increasing $\phi_{sl}$ at constant $\phi_{p}$ before saturating at
higher values of $\phi_{sl}$ (fig.~\ref{yield-stress-phi-ratio}a).  On
the other hand, the yield stress of the gel also increases
monotonically with increasing $\phi_{p}$ at constant $\phi_{sl}$
(fig.~\ref{yield-stress-phi-ratio}b). However, saturation of yield
stress values is not observed, but the data is suggestive of a rapid
rise in the yield stress beyond certain value of $\phi_{p}$.  Thirdly,
for a fixed value of $\phi_{sl}/\phi_{p}$ (simultaneous increase in
$\phi_{p}$ and $\phi_{sl}$), the yield stress shows a continuous
increase (traversing vertically upwards in
fig.~\ref{yield-stress-phi-ratio}a and b). And finally, a constant
value of yield stress ($\tau_{y}$) is observed with varying
$\phi_{sl}/\phi_{p}$ achieved either by simultaneous increase in
$\phi_{sl}$ and decrease in $\phi_{p}$ or vice-versa (traversing
horizontally in fig.~\ref{yield-stress-phi-ratio}a and b).  For all
the values of $\phi_{sl}/\phi_{p}$ lower than those reported in the
figure, a much weaker gel is obtained for which the rheometer response
is not accurate enough to obtain the flow curves. The above discussion
is highly suggestive of a non-linear dependence of yield stress on
$\phi_{sl}$ and $\phi_{p}$. We, next, carry out a scaling analysis of
the data to recover this non-linear dependence.

\begin{figure*}
  \centering \resizebox{0.8\textwidth}{!}{\includegraphics{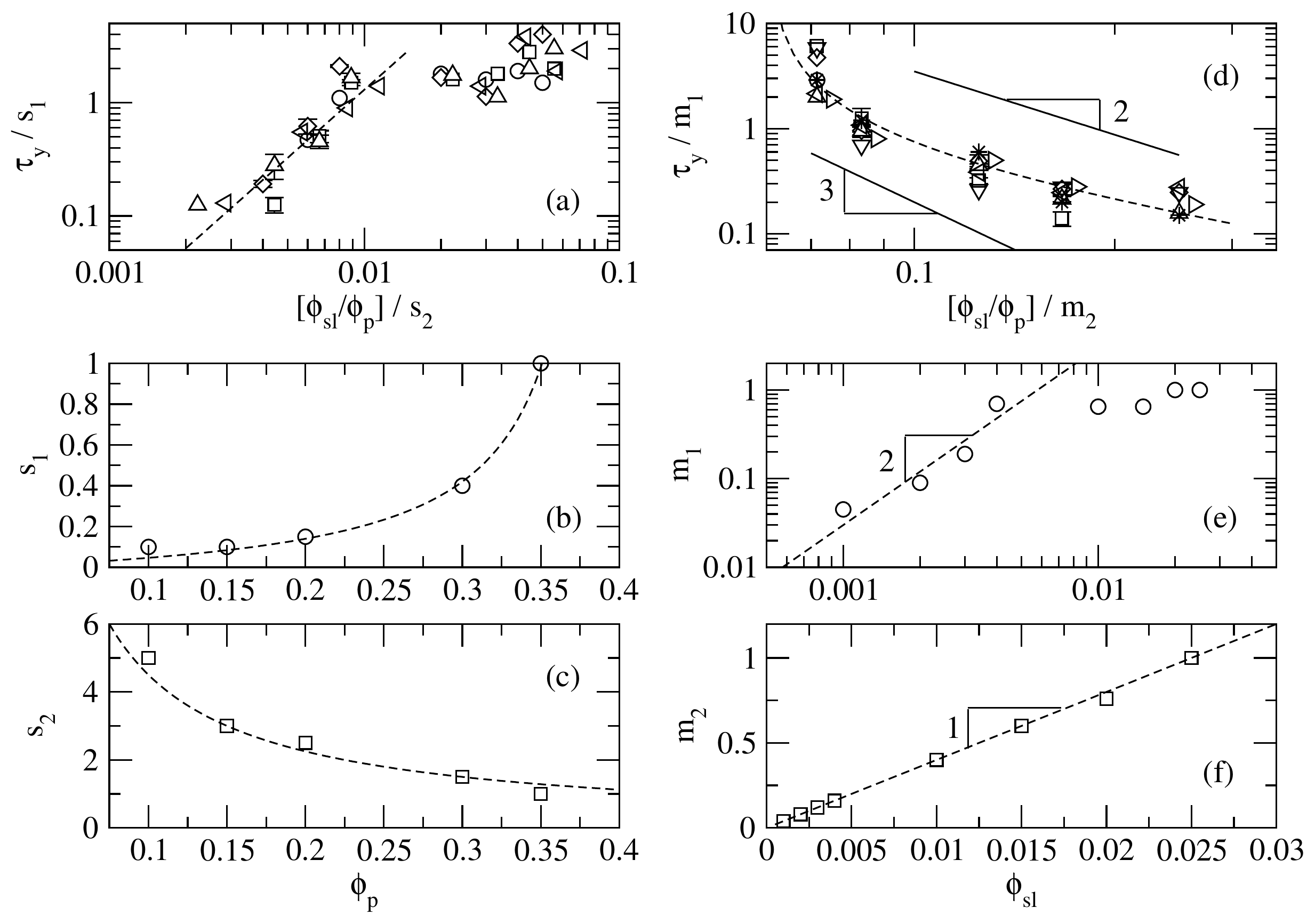}}
  \caption{Scaling analysis. Shifting of the yield stress data for
    varying values of $\phi_{sl}/\phi_{p}$ ratio obtained at (a) fixed
    values of $\phi_{p}$ and (d) fixed values of $\phi_{sl}$.  The
    error bars are shown in (a) and (d) for a few cases, the same as those
    in fig.~\ref{yield-stress-phi-ratio}. (b, c) Variation of scaling
    factors $s_{1}$ and $s_{2}$ with $\phi_{p}$, respectively. (e, f)
    Variation of scaling factor $m_{1}$ and $m_{2}$ with $\phi_{sl}$,
    respectively.  Dashed lines show specific equation fits to the
    data (see text for details). Solid lines in (d) represents
    power-law variation with exponents 2 and 3 (see text for
    more details)}
  \label{scaling}
\end{figure*}

The yield stress curves for constant $\phi_{p}$
(fig.~\ref{yield-stress-phi-ratio}a) are shifted vertically and
horizontally, using respective shift factors, $s_{1}$ and $s_{2}$ to
obtain reasonable collapse of the data as shown in
fig.~\ref{scaling}a. The variation of these shift factors with
$\phi_{p}$ is shown in fig.~\ref{scaling}b and \ref{scaling}c. Fitted
(dashed) lines to the data for shift factors yields
$s_{1} \propto (1/\phi_{p}-2.5)^{-1}$ and $s_{2} \propto
1/\phi_{p}$. Substituting these expressions for $s_{1}$ and $s_{2}$ in
the power-law model fit (of exponent 2) to the yield stress data
(shown as dashed line in fig.~\ref{scaling}a), provides the non-linear
dependence of the yield stress on $\phi_{sl}$ and $\phi_{p}$ given as
$\tau_{y} \propto \phi_{sl}^{2} (1/\phi_{p}-2.5)^{-1}$ This power-law
fit is, however, not valid at higher values of $\phi_{sl}/\phi_{p}$
for which the yield stress shows saturation and remains nearly
constant. In a similar vein, horizontal and vertical shifting of the
yield stress data for constant $\phi_{sl}$
(fig.~\ref{yield-stress-phi-ratio}b) results in the same scaling,
namely $\tau_{y} \propto \phi_{sl}^{2} (1/\phi_{p}-2.5)^{-1}$, thus
confirming the consistency of data shifting and scaling. The
variations of shift factors, $m_{1}$ and $m_{2}$, are shown in
fig.~\ref{scaling}e and \ref{scaling}f, respectively, while the final
scaling is shown as dashed line in fig.~\ref{scaling}d.  It is also
possible to fit an equation comprising of an exponential term (not
shown) to the entire range of data (increasing as well as plateau
region) from fig.~\ref{scaling}a and \ref{scaling}e instead of the
power-law fit in the increasing region of $\tau_{y}$.  Subsequent
simplification after model fitting yields an alternate expression for
the yield stress given as
$\tau_{y} \propto [1.0 - exp(-k \phi_{sl}^{2})] (1/\phi_{p}
-2.5)^{-1}$, where $k$ is a numerical constant. Assuming pendular
state gels with liquid bridging between particles, it is expected that
the yield stress would scale with $\Gamma / R$ in accordance with
eq.~(\ref{eq:yield_stress}) as also observed
previously~\citep{koos12a}. This scaling behavior, however, cannot be
verified as the particle size and liquids were not varied in the
experiments. We, next, discuss the dependence of yield stress on
secondary liquid fraction and particle volume fraction.

The shifted data in fig.~\ref{scaling}a shows a rapid increase of
yield stress followed by a very slow increase with increasing
$\phi_{sl}$ for a fixed value of $\phi_{p}$. This behavior is in
agreement with previous experimental observations for such gel
systems~\citep{kao75,koos11,koos12a,dittmann14,velankar15,bossler18}
as well as for cohesive granular systems~\citep{scheel08a}. The
increasing yield stress region is generally referred to as the
pendular regime while the region corresponding to very slow increase
is referred as the funicular regime characterised by coalescence of
liquid bridges~\citep{herming05,velankar15,bossler18}.  The yield
stress corresponds to the effective force required for rupturing
the weakest attractive (physical) bonds between particles in the gel
resulting in a shear plane along which the material yields. We
conjecture that the increase in the yield stress with increasing
secondary liquid for fixed particle content originates out of two
possibilities, viz., due to more number of particle-particle
(physical) bonds formed with added liquid and possible increase in the
volume of liquid bridge between particles (which is, however, not
possible to quantify with the available experimental set-up). The
predominant presence of the latter has been shown recently by
\citet{weis19} using tomography measurements with milimeter sized
particles and a completely wetting liquid.  The increased liquid
bridge volume should increase the capillary attractive force between
particles (see eq.~(\ref{eq:capillary-force})) provided the gel state
is in the pendular regime~\citep{willet00,herming05} which the data
seems to correspond to and will require a larger separation distance
to be achieved before breaking of the particle-particle bonds as
observed previously~\citep{willet00,mcculfor11,pitois00}. The squared
dependence on liquid content as seen in fig.~\ref{scaling}a, e or the
alternate expression comprising of an exponential term specified above
has not been observed before. The explanation will, perhaps, require a
relation between liquid bridge volume and added secondary liquid
content (specific expression for $g(\overline{V_{sl}})$ in
eq.~\ref{eq:yield_stress}), which we are unable to provide at the
moment. Such a relation exhibiting a one-third dependence of liquid
bridge volume on the ratios of the volume of secondary liquid to
particle content has been proposed previously~\citep{kao75} with the
yield stress eventually showing a power-law behavior having exponent
$1/3$ with increasing $\phi_{sl}/\phi_{p}$. For a significant increase
in the liquid content beyond that corresponding to saturation, the
yield stress has been shown to exhibit a continuous decrease (not
studied in this
work)~\citep{kao75,koos11,dittmann14,velankar15,bossler18}.

An increase in the yield stress is also observed with increasing
$\phi_{p}$ for constant $\phi_{sl}$ as seen from the shifted data in
fig.~\ref{scaling}d. However, the behavior suggests a rapid increase
occurring around $\phi_{p} = 0.4$ as obtained from the fitted
equation.  Increase in the number of particles present in the gel
corresponds to increased number of attractive physical bonds formed
between particles, thereby leading to higher yield stress. However,
given that liquid content is maintained the same, larger number of
bonds are, perhaps, formed either by more optimal usage of the
existing liquid or due to excess liquid droplets in the system which
were never used in the network, the latter case being more probable
(see fig.~\ref{liquid-bridge-distribution}).  The effect of increased
crowding at larger $\phi_{p}$ supposedly supplements the capillary
bridging induced particle connectivity making the gel that much harder
to shear thereby showing a faster increase in the yield stress. The
precise value of $0.4$ is, however, the result of the fitting to the
experimental data over the limited range of $\phi_{p}$ values studied
over here. Overall, the dependence of the yield stress on particle
fraction is stronger compared with that on the secondary liquid
content.

Figure~\ref{scaling}d also shows a power-law dependence of $\tau_{y}$
on $\phi_{p}$ with two different exponents. The power-law variation
with exponent $2$ agrees with previous predictions
~\citep{kao75,pietsch67}. The power-law fit with exponent $3.3$ as
observed by \citet{velankar15} and that with an exponent value around
$4$, as recently reported by \citet{bossler18} however seems to agree
only for the higher range of $\phi_{p}$. The observed discrepancy with
our experimental results over the entire range of $\phi_{p}$ studied
can be attributed partly to the differences in the type of flow curves
observed in different studies and partly to the very large particles
used (hence weaker gels) for our system. An even higher value of
power-law exponent ($4.6$) has been observed for a gel considered to
be arising out of hydrogen bonding~\citep{cavalier02}, but later on
shown to be actually arising due to capillary bridging~\citep{koos11}.
We also note recent theoretical and simulation study for particles
with attractive forces which have shown such power-law dependence with
exponents $3.5$ and $3.9$, respectively, but for a compressive yield
stress~\citep{roy16c}.

\subsection{Gel microstructure}
\label{micro}

\begin{figure*}
  \centering \resizebox{0.8\textwidth}{!}{\includegraphics{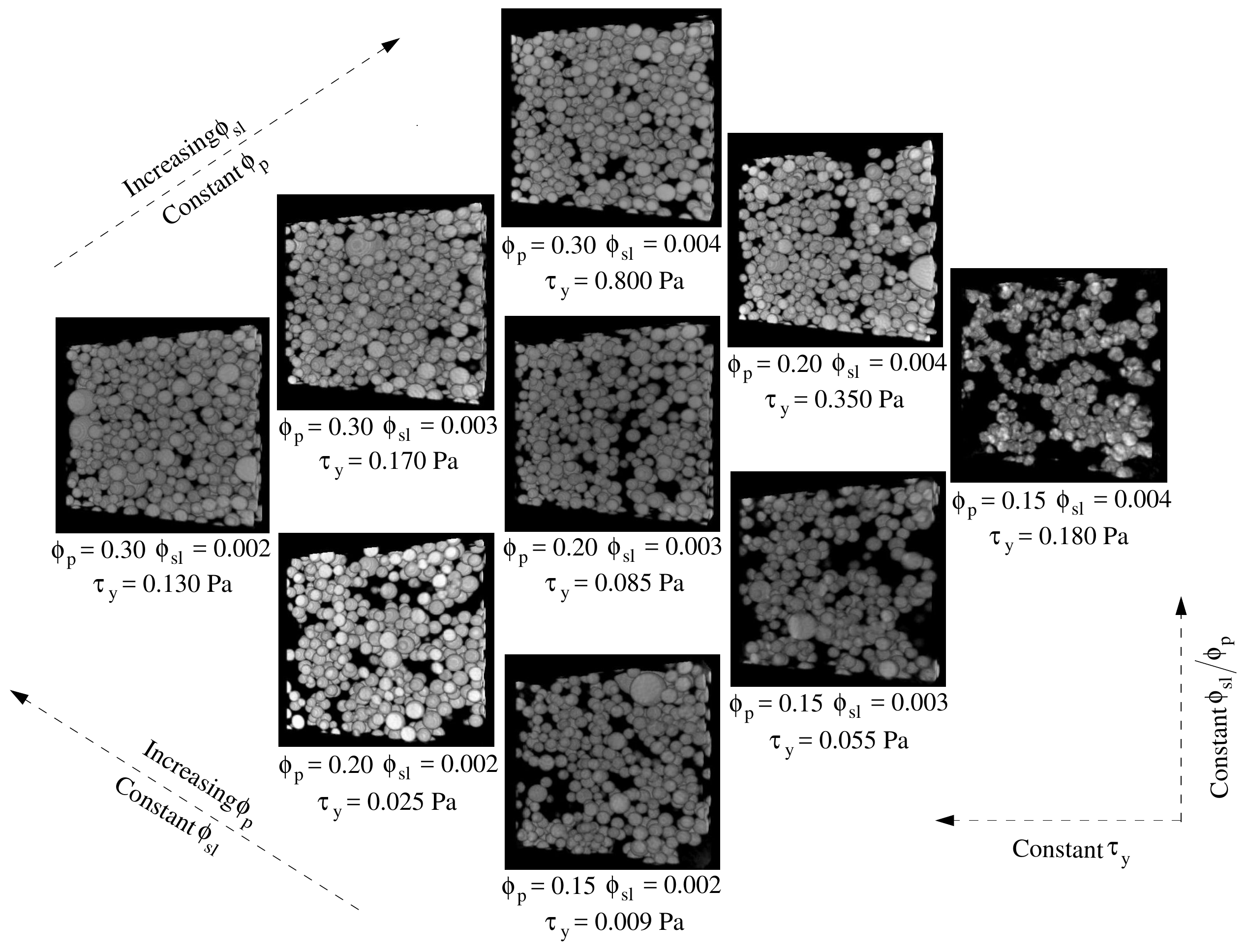}}
  \caption{X-ray tomography images exhibiting the microstructure
    obtained for varying values $\phi_{sl}$ and $\phi_{p}$. The yield
    stress value corresponding to each microstructure are also
    mentioned in the figure (see text for more details about
    tomography measurement)}
  \label{gel-microstructure}
\end{figure*}

The three dimensional reconstructed images (from a perspective view
angle) for a few cases, as obtained from X-ray tomography technique,
are shown in the fig.~\ref{gel-microstructure}. A similar diagram, but
only schematically, was proposed previously~\citep{velankar15} to
emphasise, qualitatively, the observed bulk rheology behavior. All
images represent the values of particle to liquid volume ratios over
which the yield stress varies and has not saturated to a constant
value. Each image represents a $3$D slice of size
$20 \times 20 \times 10$ particle diameters, while the movies
corresponding to each of these images (of a larger region
$20 \times 20 \times 20$ particle diameters) are available as
supplementary material. A thinner slice is included over here for ease
in visualisation of the three dimensional structure.  As seen from the
figure, each image clearly depicts a particular microstructure for
specified $\phi_{sl}$ and $\phi_{p}$ corresponding to a different
yield stress ($\tau_{y}$). The variation of $\phi_{sl}$, $\phi_{p}$,
their ratio and $\tau_{y}$ is clearly shown in the figure and is in
accordance with the data shown in fig.~\ref{yield-stress-phi-ratio}.

\begin{figure*}
  \centering \resizebox{0.8\textwidth}{!}{\includegraphics{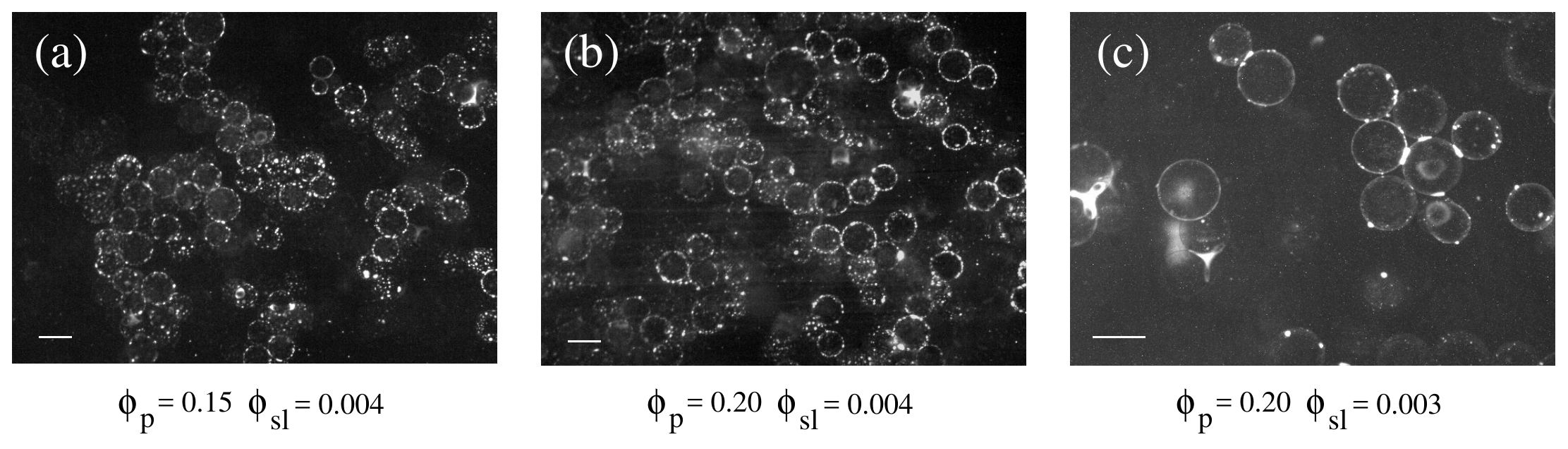}}
  \caption{Two dimensional images providing visual
    presence/distribution of the secondary liquid within the gel
    microstructure. Each image is obtained by a laser sheet
    illuminating a plane away from the walls (in the bulk) and
    fluorescing the dyed secondary liquid (see text for imaging
    details). Images in (a) and (b) are representative of the three
    dimensional matrix obtained by sweeping the laser sheet in the
    sample. (c) Image showing a much closer view depicting the liquid
    bridges formed between the particles. The scale bar shown at the
    bottom left in each figure corresponds to $350$ microns.}
  \label{liquid-bridge-distribution}
\end{figure*}

The detailed view of the distribution of the secondary liquid (TG-1P
mixture) in the gel matrix for three specific cases is shown in
fig.~\ref{liquid-bridge-distribution}. A thin laser sheet illuminates
a plane in the bulk (about $10-15$ particle diameters from end walls)
of the sample placed in a transparent walled rectangular cell and
fluorescing the dye present in the secondary liquid to generate the
two dimensional images shown in the figure. The un-dyed bulk liquid
and the particles are not visible due to very good refractive index
matching achieved. The particle presence is, however, perceived from
the bright circular rings which represent secondary liquid droplets
accumulating around the particle. A very bright puddle represents a
very large drop of secondary liquid, perhaps not effectively broken
into small micro-droplets during sample preparation. Note that not
every drop of secondary liquid participates in particle
bonding/connectivity in the sample. However, those participating do
form liquid bridges connecting two or more particles as exhibited in
the closer view shown in fig.~\ref{liquid-bridge-distribution}c.  In
this scenario, the exact amount of the secondary liquid involved in
liquid bridging cannot be estimated. Nevertheless, increased content
of secondary liquid, measured as $\phi_{sl}$, does show a direct
correlation with increasing and saturating yield stress.

The image matrix shown in fig.~\ref{gel-microstructure} suggests four
key microstructural behaviours. Increasing value of $\phi_{sl}$ for
fixed $\phi_{p}$ seemingly increases the local porosity, but only
slightly. This represents slight rearrangements between particles to
form more bonds (more local compactness) while maintaining the base
sample spanning structure, responsible for gel behavior. An increase
in the porosity value with increasing value of $\phi_{sl}$ followed by
saturation at very high values of $\phi_{sl}$ (akin to yield stress
behavior) has been observed previously, accompanied by qualitative
differences in the pore size distribution for different values of
$\phi_{sl}$~\citep{dittmann14}. Now, keeping the value of $\phi_{sl}$
fixed while increasing the value of $\phi_{p}$ decreases the porosity
due to crowding by additional particles which occupy the void space
and form more bonds. The yield stress in both cases increases
significantly. Varying both quantities simultaneously provides
interesting behaviours. A proportionate increase in $\phi_{sl}$ and
$\phi_{p}$ to maintain a constant ratio causes increased crowding
(more particles, more liquid, more bonding). The yield stress, quite
obviously, increases significantly in this case.  Finally, increasing
$\phi_{sl}$ and decreasing $\phi_{p}$ simultaneously or vice-versa,
respectively, increases or decreases microstructural porosity, but
interestingly maintains a constant yield stress for the gel system.
The yield stress in the latter is through increased crowding (bonding)
and in the former case is, perhaps, due to increased liquid bridge
volume and slight rearrangement in the microstructure. To summarise, a
wide range of microstructures are realizable by fine tuning
$\phi_{sl}$ and $\phi_{p}$ to achieve the desired strength (yield
stress) for the gel system. We next try to explain the observed
microstructural and yield stress behavior using some quantitative
measurements.

The images obtained from X-ray tomography measurements were analysed
using centroid algorithm in IDL (Interactive Data Language), after
appropriate filtering and contrast adjustment, to obtain the particle
centroids and radius to a pixel accuracy. Given the noise/distortion
in the images, about $85\%$ of particles were detected in each
sample. This data was then used to determine the particle connectivity
in the gel sample. Two particles were considered to be bonded to each
other (due to liquid bridge) if the geometric distance $d_{c}$ between
their individual centroids is less than $1.15$ $(D_{1}+D_{2})/2$,
where $D_{1}$ and $D_{2}$ are the diameters of two particles. The
value $1.15$ was fixed to account for the possible errors in centroid
and particle radius determination and was arrived at after analysing
image sets containing particles of known centroids and radii created
manually. The number of bonds associated with every detected particle
was obtained using above arguments for each of the nine cases shown in
fig.~\ref{gel-microstructure}. The data, averaged over $2-3$
experimental runs (tomography sets) for each case, was then used to
calculate the distribution $P(n_{bond}$) where, $n_{bond}$ is the number of
bonds associated with a particle and $P(n_{bond})$ is the number of
particles in the system exhibiting $n_{bond}$ number of
bonds. The summation of the values of $P(n_{bond})$ over
  all values of $n_{bond}$ then corresponds to the average number of
  particles detected in an image. In the following, the qualitative
behavior of the distributions is discussed for three parametric
variations, viz. increase in $\phi_{sl}$ for fixed values of
$\phi_{p}$, increase in $\phi_{p}$ for fixed values of $\phi_{sl}$ and
increase in both, $\phi_{sl}$ and $\phi_{p}$, while keeping the ratio
$\phi_{sl}/\phi_{p}$ constant.

The value of $P (n_{bond})$ for $n_{bond} = 0$ represents
 number of particles with no connectivity. This number varies from
  around $180$ (about $10-12$\% for total particles identified) for
  the smallest liquid and particle fraction to around $70-80$ (about
  $3.5$\% of total particles identified) for the largest liquid and
  particle contents studied. The number arises due to the (i)
  possibility of individual particles remaining suspended in the gel
  without connecting to any cluster (ii) possible errors in detecting
  connected particles and (iii) the inability of the image analysis
  algorithm to detect all the particles in the system.  The relative
  contribution of these three effects is, however, not possible to
  quantify.

\begin{figure*}
  \centering \resizebox{0.8\textwidth}{!}{\includegraphics{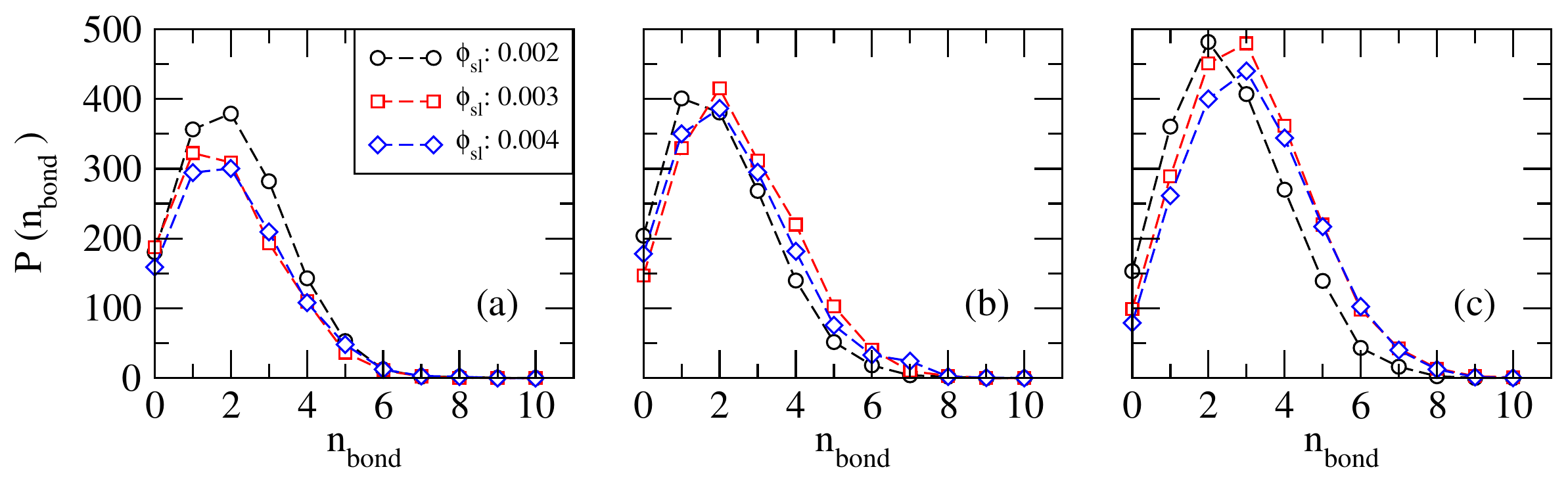}}
  \caption{Variation of the distribution of the
      number of particles having $n_{bond}$ number of bonds with
      $\phi_{sl}$ for (a) $\phi_{p} = 0.15$ (b) $\phi_{p} = 0.20$ and
      (c) $\phi_{p} = 0.30$.}
  \label{cluster-distribution-phisl}
\end{figure*}

\begin{figure*}
  \centering \resizebox{0.8\textwidth}{!}{\includegraphics{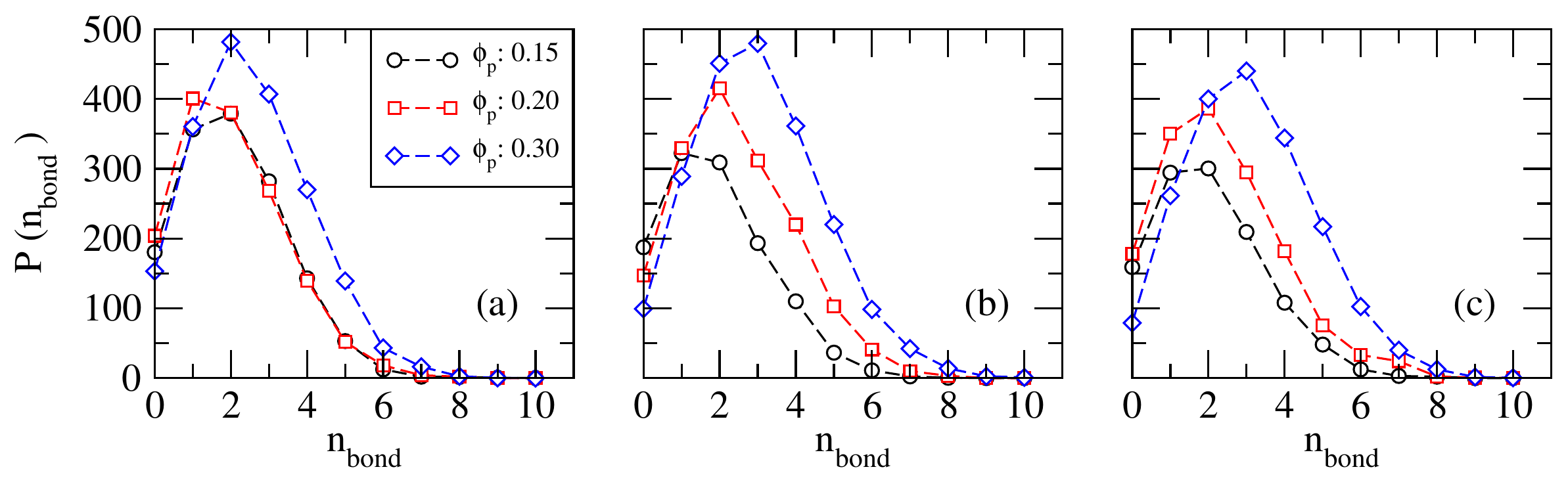}}
  \caption{Variation of the distribution of the
      number of particles having $n_{bond}$ number of bonds with
      $\phi_{p}$ for (a) $\phi_{sl} = 0.002$ (b) $\phi_{sl} = 0.003$
      and (c) $\phi_{sl} = 0.004$.}
  \label{cluster-distribution-phip}
\end{figure*}

\begin{figure*}
  \centering \resizebox{0.8\textwidth}{!}{\includegraphics{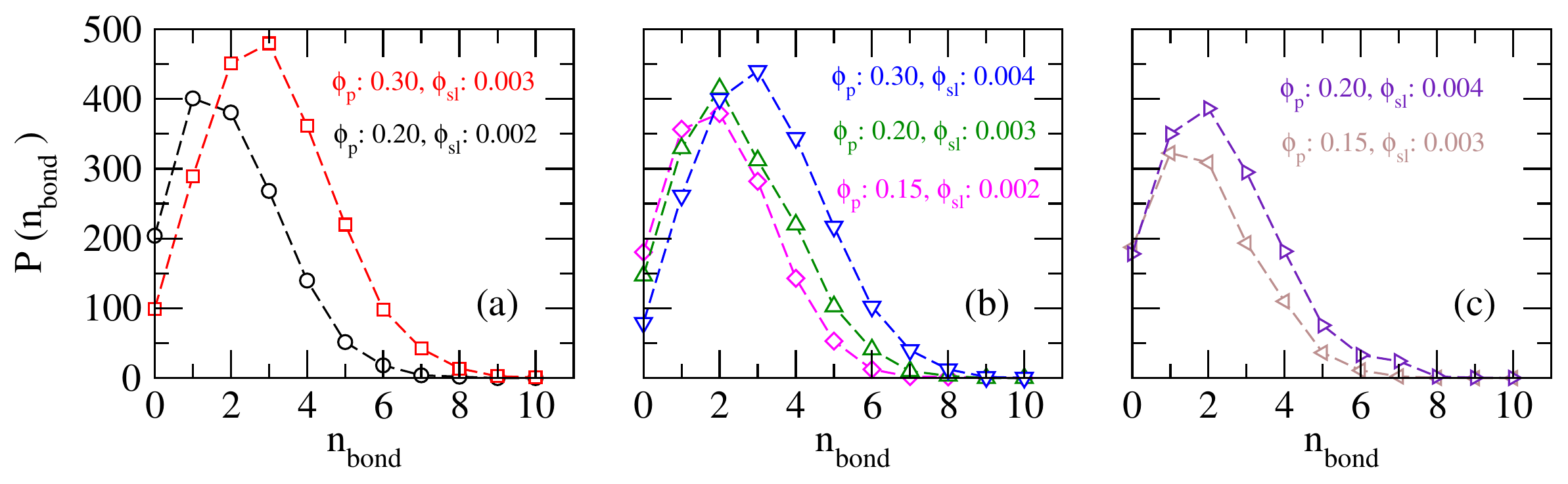}}
  \caption{Variation of the distribution of the
      number of particles having $n_{bond}$ number of bonds for (a)
      $\phi_{sl}/\phi_{p} = 0.01$ (b)
      $\phi_{p}/\phi_{p} \approx 0.013$ and (c)
      $\phi_{sl}/\phi_{p} = 0.02$.}
  \label{cluster-distribution-ratio}
\end{figure*}

We next look at the effect of increasing content of secondary liquid
for fixed value of $\phi_{p}$. This corresponds to an increase in the
ratio $\phi_{sl}/\phi_{p}$ and represents a right-upward travel in
fig.~\ref{gel-microstructure}. The distribution, $P(n_{bond})$, is shown
in fig.~\ref{cluster-distribution-phisl}.  As discussed earlier, the
increased liquid content does visually show a slight change in
porosity as noted in fig.~\ref{gel-microstructure} while preserving
the sample spanning network resulting in a gel-like behavior. This
slight change in porosity is not reflected appreciably in the
distribution.  A slight rightward shift at highest particle fraction
is observed in (fig.~\ref{cluster-distribution-phisl}c), which is most
likely, due to larger number of particles available for possible
rearrangement of the microstructure. The yield stress, however, shows
an appreciable change, from $0.009$ Pa to $0.18$ Pa in case (a), from
$0.025$ Pa to $0.35$ Pa in case (b) and from $0.13$ Pa to $0.8$ Pa in
case (c).  The increase in the yield stress can then be attributed
primarily to the increase in the liquid bridge volume (discussed
earlier in conjunction with eq.~(\ref{eq:capillary-force})) and slight
contribution due to increased number of bonds per particle at highest
particle fraction~\citep{weis19}. The exact dependence of liquid
bridge volume on $\phi_{sl}$ as shown
previously~\citep{willet00,herming05,pitois00,mcculfor11}, but for
contact angles much smaller than $60$ degrees observed over here, is,
however, not possible to quantify.

The distribution, $P(n_{bond})$, for increasing values of $\phi_{p}$ at
fixed values of $\phi_{sl}$ is shown in
fig.~\ref{cluster-distribution-phip}. The variation represents
decreasing value of the ratio $\phi_{sl}/\phi_{p}$ and corresponds to
left-upward travel in fig.~\ref{gel-microstructure}. Unlike previous
case, the effect of particle fraction is observed quite
distinctly. The distribution show a rightward shift with increasing
particle content as well as increasing value at the peak. It can be
expected that increasing the particle content ($\phi_{p}$), while
keeping the liquid content ($\phi_{sl}$) constant will allow for more
number of particles available for bond formation. Given the fact that
not all the secondary liquid is utilised in bond formation (see
fig.~\ref{liquid-bridge-distribution}), increasing particle content
will lead to more effective use of the available liquid leading to
increased number of bond formation and consequently rightward shift in
the distribution as well as an increase in the maximum value.  The
yield stress shows a monotonic increase, from $0.009$ Pa to $0.13$Pa
in case (a), from $0.055$ Pa to $0.17$ Pa in case (b) and from $0.18$
Pa to $0.8$ Pa in case (c). This increase in the yield stress can be
attributed primarily to the increase in the number of bonds per
particle with increasing $\phi_{p}$ given that $\phi_{sl}$ is
maintained constant in each case.

The simultaneous increase in $\phi_{sl}$ as well as $\phi_{p}$ so as
to maintain the ratio $\phi_{sl}/\phi_{p}$ constant corresponds to
vertically upward traversing in the images shown in
fig.~\ref{gel-microstructure}.  The distribution for these cases is
shown in fig.~\ref{cluster-distribution-ratio}. The distribution shows
a rightward shift and an increased peak value in each case.
Simultaneous increase of both the entities leads to (i) large number
of bonds formed per particle due to increased particle and liquid
availability resulting in a qualitative change in the distribution and
(ii) possible increase of liquid bridge volume due to increased liquid
content. The increase in the yield stress, from $0.025$ Pa to $0.17$
Pa in case (a), from $0.009$ Pa to $0.8$ Pa in case (b) and from
$0.044$ Pa to $0.35$ Pa in case (c), can be attributed to
contributions from both, increased number of bonds per particle and
possible increase in liquid bridge volume~\citep{weis19}.

The overall behavior of the microstructure discussed above suggests
that the distribution, $P(n_{bond})$, is primarily dependent on
$\phi_{p}$ and only marginally on $\phi_{sl}$, whereas the yield
stress shows significant dependence on $\phi_{sl}$ as well as
$\phi_{p}$. Further, the change in the number of bonds per particle
does not seem to be the only factor that correlates with the changes
in the yield stress. The data suggests that the changes in the yield
stress can also occur due to possible change in the liquid bridge
volume that cannot be quantified appropriately. Further, it is not
possible to correlate the extent of the increase in the yield stress
in every case with the distribution curves.

\section{Conclusions}
\label{concl}
In this study, we have experimentally investigated the flow and
structure of a capillary attractive force initiated gel using rheology
and tomography techniques. The experiments were carried out for a
range of particle volume fraction and varying concentration of
secondary liquid which induces the necessary capillary attractive
force. The rheology behavior predominantly comprises of a yield stress
with a linear behavior of shear stress at high shear rates.  The yield
stress extracted from the flow curves shows, hitherto unobserved,
highly non-linear dependence on the secondary liquid fraction and
particle concentration. The dependence on the particle fraction is,
however, stronger compared to that on the secondary liquid. A
power-law dependence, of exponent $2$, is observed for secondary
liquid fraction for small enough values of $\phi_{sl}/\phi_{p}$, while
a rapid increase is observed with increased particle volume
fraction. Microstructural details of the gel have been extracted using
direct visualisation in three dimensions through X-ray tomography
followed by image analysis to obtain particle locations
(centroids). The local porosity of the gel is observed to increase
with increasing secondary liquid fraction due to more localised
compaction by particle rearrangements, while it decreases with
increasing particle fraction due to crowding, the yield stress
increasing in both the cases. In the former case, the increase in
yield stress can be attributed primarily due to increase in the liquid
bridge volume and a very slight change in the number of bonds per
particle. In the latter case, the yield stress seems to increase
primarily due to increase in the number of bonds between particles,
each contributing to capillary attractive force which needs to be
overcome to initiate yielding. The above behavior is reflected
reasonably well through the distributions of particle-particle bonding
(or coordination number) obtained from image analysis.

The observed rheology should provide a strong impetus towards modeling
of the yield stress behavior for such gel systems. It is not quite
clear whether the differences in the behavior are only due to the
larger particle sizes (weaker gels) employed in this work compared to
all other studies. Theoretical constitutive modeling will then, have
to be formulated to encompass the qualitative change in the observed
behavior across a range of particle sizes. Equally interesting and
challenging would be to investigate the behavior of such gel systems
for (i) spherical particles with heterogeneity in the wettability
across particle surface (Janus particles) (ii) binary mixture of
particles based on their wettability, one wettable with respect to
primary liquid while other with respect to secondary liquid and (iii)
particles with aspherical or rod-like shapes. Understanding the
mechanism of gelation, structural formation, strength of the gels for
such systems which are quite different than those being studied till
now and which also are of much more practical relevance should form
the scope of the future research work for this exciting gel forming
system.

\section{\label{suppl}Supplementary material}

Movies corresponding to all the images shown in
fig.~\ref{gel-microstructure} are provided as supplementary material
for ready reference. Each movie is obtained using X-ray tomography
imaging over a region ($20 \times 20 \times 20$ particle diameters) by
rotating the sample over $360$ degrees.

\begin{acknowledgement}
  We thank Mahesh Tirumkudulu for insightful suggestions which helped
  us to improve upon the manuscript. We are also very thankful to
  Mr. Arun Torris for the assistance provided in tomography imaging,
  Mayuresh Kulkarni for assistance in laser imaging, Arun Banpurkar
  for providing help in measuring the surface tension of all liquids
  and Shankar Ghosh for assistance in refractive index measurements of
  all the materials. The financial support from Science \& Engineering
  Research Board, India (Grant No. SB/S3/CE/017/2015) is gratefully
  acknowledged.
\end{acknowledgement}

\appendix
\section{Appendix}
\subsection{Rheology dependence on shear gap}
\label{gap-depen}

\begin{figure}
  \centering \resizebox{0.4\textwidth}{!}{\includegraphics{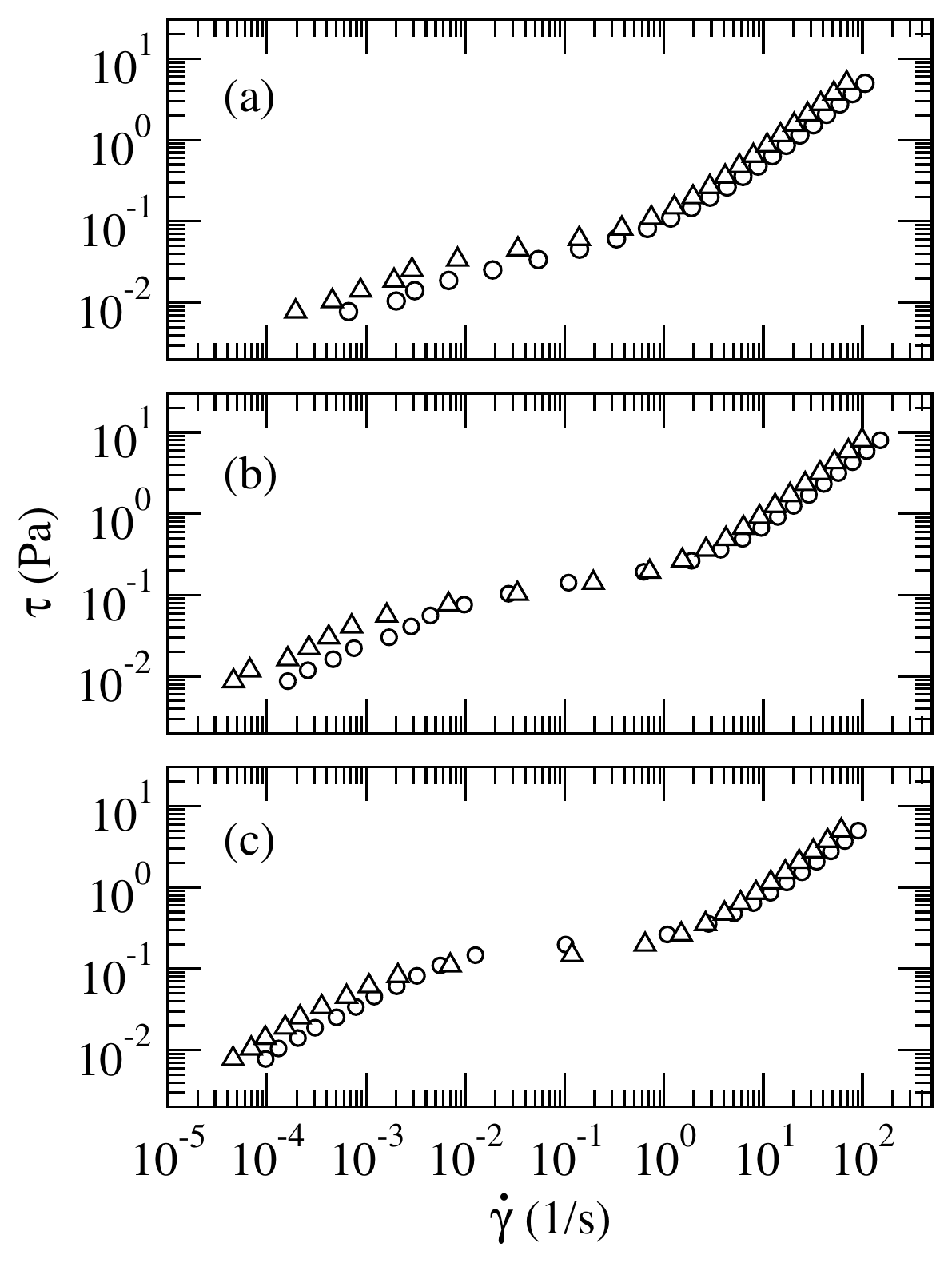}}
  \caption{Effect of shear gap on stress ramp rheology behavior. (a)
    $\phi_{p}: 0.15, \phi_{sl}: 0.002$ (b)
    $\phi_{p}: 0.20, \phi_{sl}: 0.003$ and (c)
    $\phi_{p}: 0.30, \phi_{sl}: 0.002$. Circles and triangles,
    respectively, represent the data acquired for shear gap of $3$ mm
    ($\sim 10$ particle diameters) and $6$ mm ($\sim 20$ particle
    diameters).}
  \label{rheology-shear-gap}
\end{figure}

Stress ramp measurements were performed for three combinations of
particle fraction ($\phi_{p}$) and secondary liquid fraction
($\phi_{sl}$) in concentric cylinder (cup-bob) geometry with outer
cylinder (cup) of two different radii, viz. $11.33$ mm and $14.33$
mm. This realizes two shear gaps $3$ mm and $6$ mm since the inner
cylinder (bob) is of radius of $8.33$ mm. Given the particle mean
diameter of $D = 350 \pm 50$ microns, the two shearing gaps then
correspond to approximately $8-9$ and $16-18$ particle diameters. The
variation of stress versus shear rate for two shear gaps and different
particle-liquid fraction combination is shown in
fig.~\ref{rheology-shear-gap}. The consistently smaller values of
stress for the smaller gap compared to those for the larger gap (by a
factor or $50$\% or less) over the entire shear rate range suggests
the presence of a small amount of slip. Further, the effect of slip,
even if small, seems to be more pronounced for the data below
yielding. At the same time, the near similar behavior for both gaps
indicate that the overall results shown for $3$ mm shear gap in the
main text and the conclusions drawn from them are not affected
qualitatively by the presence of a small amount of slip. Morevoer, the
values of yield stress which is the main focus of this work remain
nearly the same for both the shear gaps.

\subsection{Stress strain relationship}
\label{stress-strain}
\begin{figure}
  \centering \resizebox{0.4\textwidth}{!}{\includegraphics{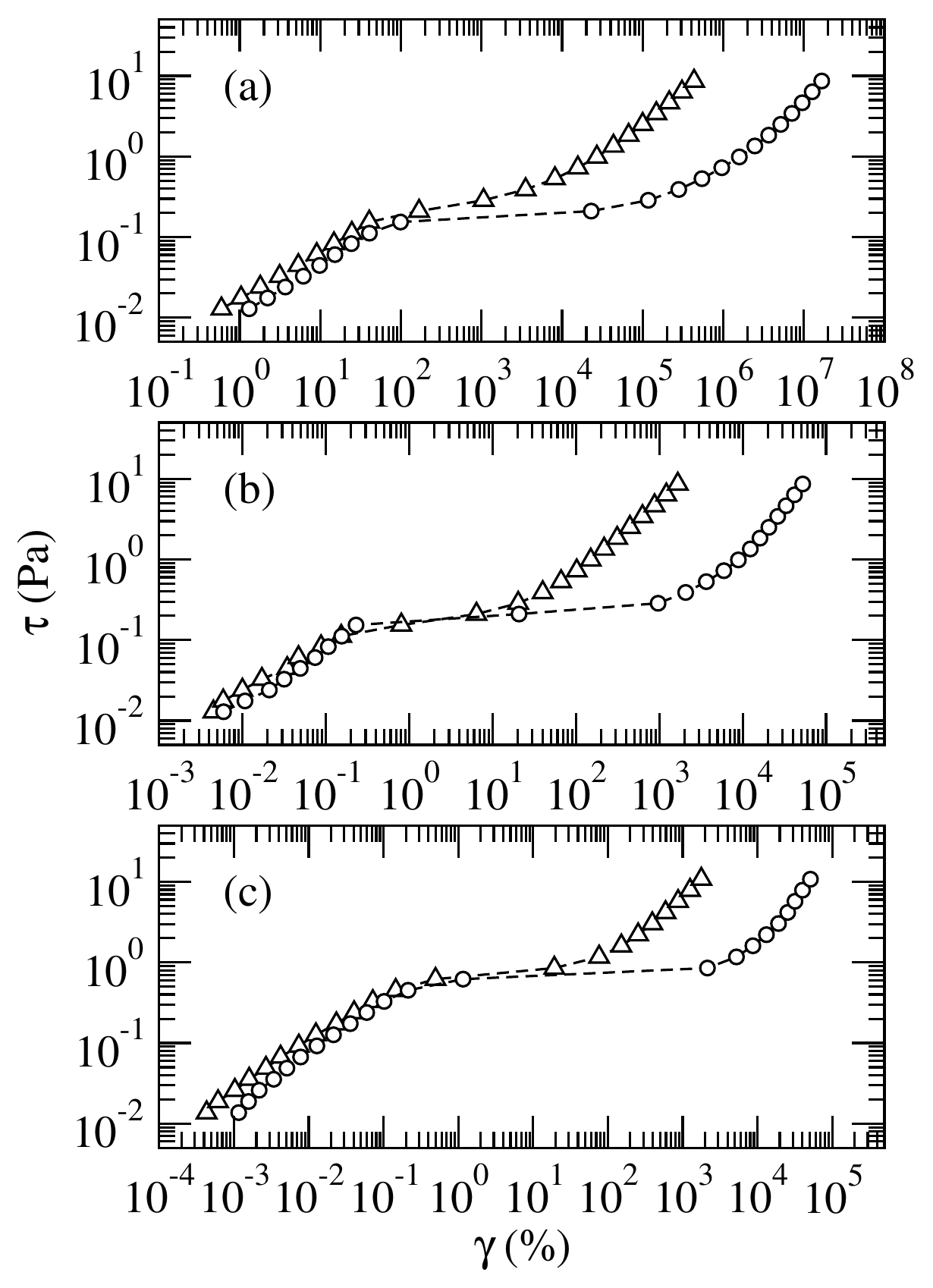}}
  \caption{Variation of shear stress with strain across the entire
    stress ramp change. (a) $\phi_{p}: 0.20, \phi_{sl}: 0.003$ (b)
    $\phi_{p}: 0.20, \phi_{sl}: 0.004$ and (c)
    $\phi_{p}: 0.30, \phi_{sl}: 0.002$. Triangles and circles,
    respectively, represent the data acquired for acquisition time
    $t_{aq} = 10$ seconds and $t_{aq} = 100$ seconds.}
  \label{stress-strain-elasticity}
\end{figure}

Stress ramp rheology experiments were performed for three combinations
of $\phi_{p}$ and $\phi_{sl}$ and for two different acquisition times,
viz. $t_{aq} = 10$ seconds and $t_{aq} = 100$ seconds. The variation
of the applied stress and corresponding recorded strain experienced by
the sample at the end of the acquisition time is shown in
fig.~\ref{stress-strain-elasticity}. The stress-strain data for both
the acquisition time shows reasonable superimposition below the
yielding, which points towards a possible elastic behavior, though the
data shown in fig.~\ref{strain-evolution-plasticity} suggests plastic
deformation. Further, the curves, above the yield point, are
distinctly different for both the acquisition times suggestive of
predominantly plastic deformation of the sample.  The yielding
behavior, however, remains unchanged by the acquisition times as seen
from the near identical yield stress for both cases. The exact
mechanism of this apparent plastic as well as possible elastic
deformation below yielding is not clear and will require flow imaging
in the limit of low shear rates not within the scope of this work.
 
\subsection{Time dependent strain evolution}
\label{strain-evol}
\begin{figure}
  \centering \resizebox{0.4\textwidth}{!}{\includegraphics{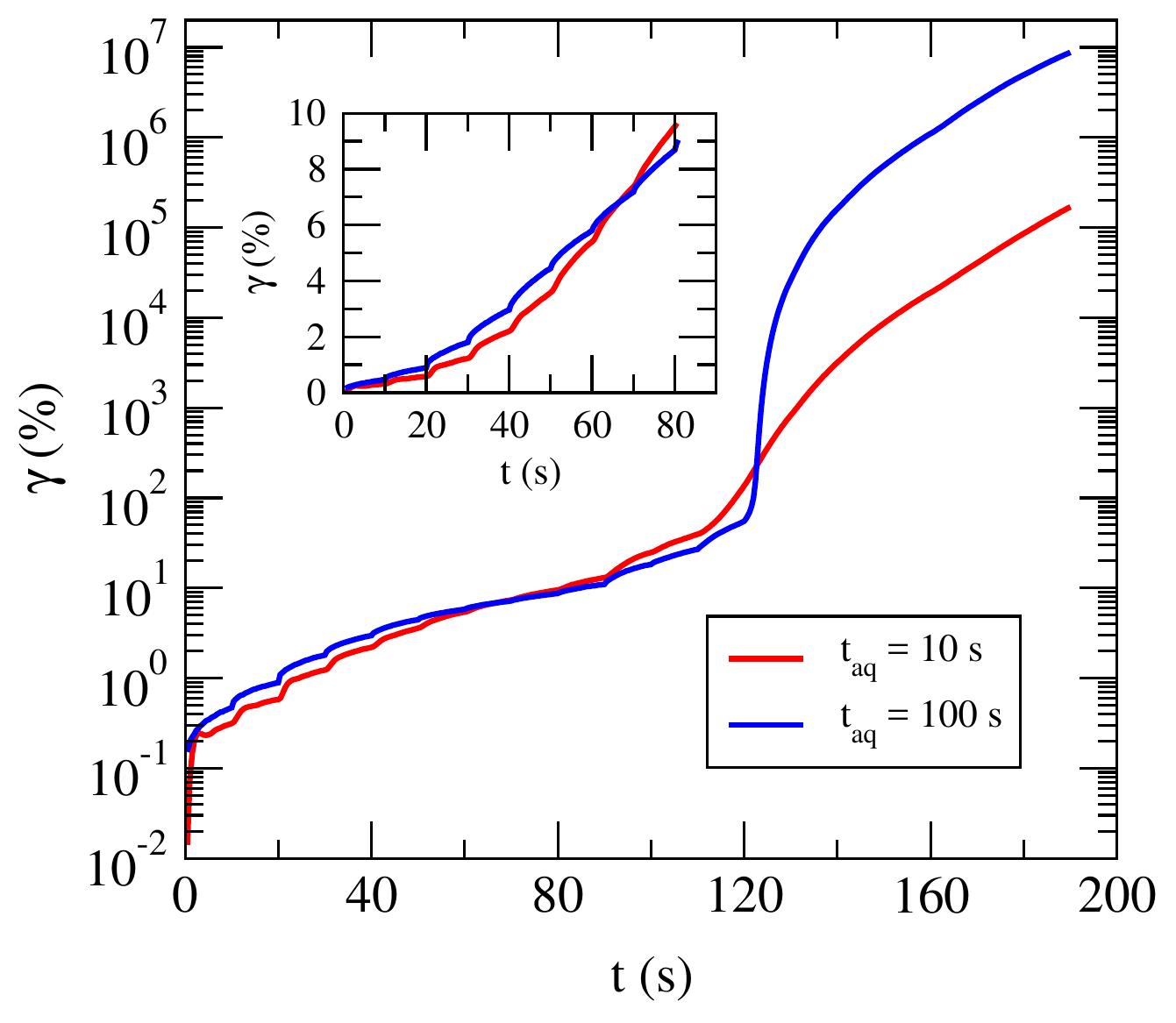}}
  \caption{Cumulative evolution of strain with time at two different
    acquisition time ($t_{aq}$) during the step-stress ramp experiment
    for $\phi_{p}:0.2$ and $\phi_{sl}: 0.003$. The applied stress
    values increase in a step-wise manner from $0.00865$ Pa to $8.65$
    Pa.  Inset: Data at early times ($t < 80 s$) shown on a linear
    scale.}
  \label{strain-evolution-plasticity}
\end{figure}

The cumulative evolution of strain ($\gamma$) with time ($t$) during a
step-stress ramp experiment is shown in
fig.~\ref{strain-evolution-plasticity} for two different acquisition
times ($t_{aq}$). Different gel samples, albeit with same
$\phi_{sl} = 0.003$ and $\phi_{p} = 0.2$, are used to obtain data for
each acquisition time. Stress was increased in step-wise manner from
$0.00865$ Pa to $8.65$ Pa and the time interval between two
consecutive stress levels was equal to the acquisition time.  The data
acquired with $t_{aq} = 100$ s was re-scaled by dividing time ($t$) by
a factor of $10$. This is done to enable direct comparison of the
strain evolution data between the experiments with $t_{aq} = 10$ s and
$t_{aq} = 100$ s. In both acquisition time experiments, the data shows
a yield event in which the strain increases rapidly beyond a certain
stress (or strain) level. Before the yield event the strain values are
small suggestive of a solid-like behavior. However, for each level of
applied stress the strain jumps slightly and then increases more
slowly with time, which suggests that the material in its pre-yield
stage actually creeps slowly. Thus, the pre-yield state is not truly
elastic, but comprises plastic deformation to certain extent, perhaps,
arising due to rearrangement of particle-particle bonds, i.e. breakage
of bonds and reforming with different particles resulting in a local
rearrangements. Because of the creep flow, the strain values with
$t_{aq} =100$ s are always higher than those with $t_{aq} = 10$ s.
Interestingly, the sample with $t_{aq} = 10$ s tends to yield (i.e.,
the strain starts increasing rapidly) at a slightly lower stress value
($0.162$ Pa) than the sample with $t_{aq} = 100$ s which tends to
yield at $0.216$ Pa.  The yield event for the sample with
$t_{aq} =100$ s is also more distinct than that for the sample with
$t_{aq} = 10$ s. These differences are likely because the sample with
$t_{aq} = 100$ s is also ageing to a greater extent during the
acquisition time compared to the sample with $t_{aq} = 10$ s. Ageing
probably results in stronger network, which resists yielding and also
breaks more distinctly. From the behavior described above, the exact
mechanism of yielding is not entirely clear. It seems that the
solid-like material in the pre-yield state actually creeps slowly
until it yields. However, the yield strains for both cases are not
identical (about $48$\% strain with $t_{aq} = 10$ and about $58$\%
with $t_{aq} = 100$ s).  Further investigations of the rheology and
yielding behavior of the suspension are needed which will require
rheo-visualization, not within the scope of the present work.

\subsection{Refractive index and density matching details}
\label{ri-den-match}
\begin{figure}
  \centering \resizebox{0.4\textwidth}{!}{\includegraphics{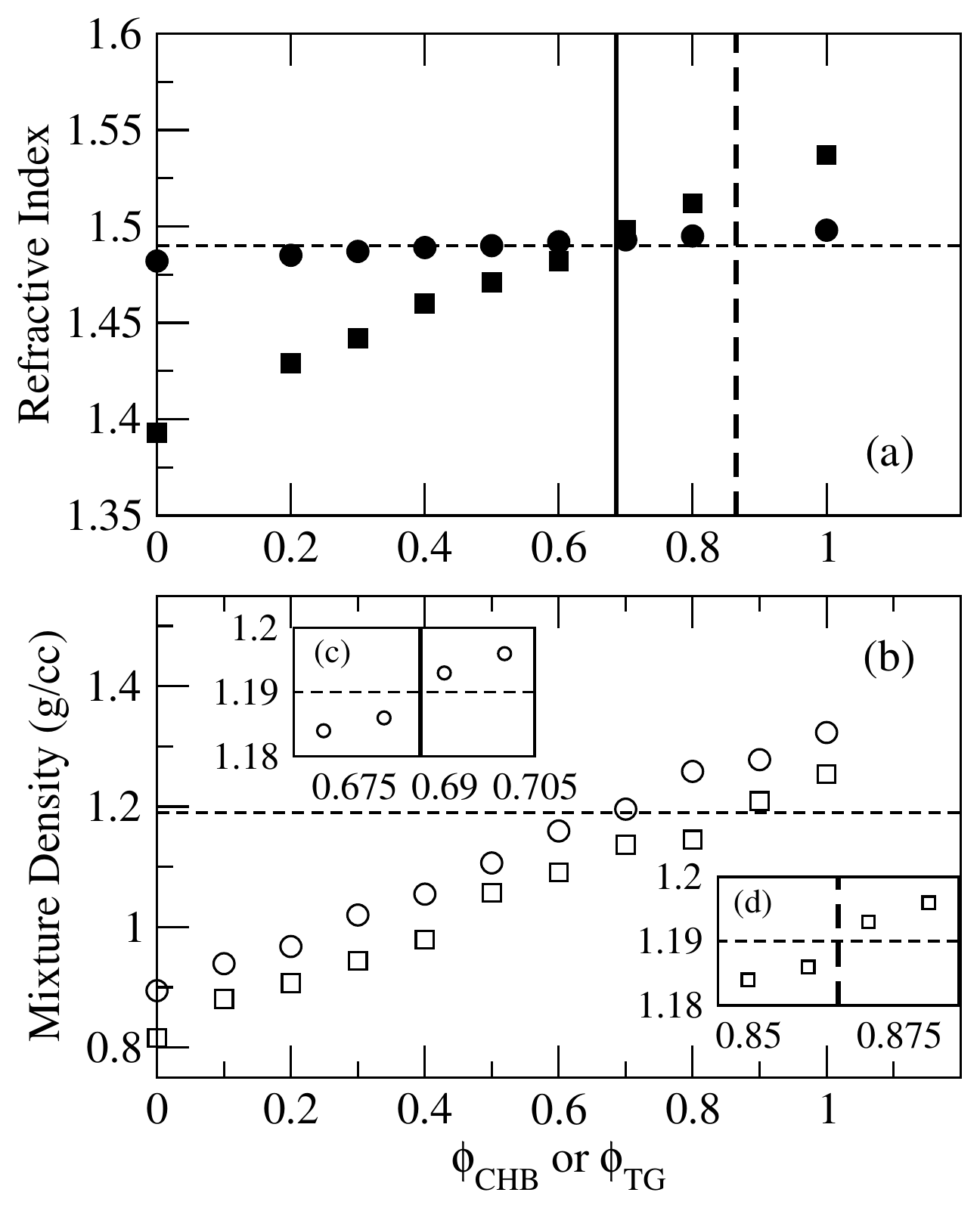}}
  \caption{Variation of the measured values of (a) refractive index
    and (b) density of the liquid mixture with volume fraction of one
    of its constituents (TG or CHB).  Circles represent data for
    $\phi_{CHB}$ in its mixture with DEC. Squares represent values for
    $\phi_{TG}$ in its mixture with 1P. The horizontal dashed line in
    (a) and (b), respectively, represents the reported values of
    refractive index and density for PMMA. Inset: An expanded view of
    the variation of mixture density for small changes to (c)
    $\phi_{CHB}$ and (d) $\phi_{TG}$ in their respective mixtures. The
    solid vertical line in (c) represents the value of $\phi_{CHB}$
    desired for accurate density matching, while the solid vertical
    line in (a) provides the closeness of refractive index matching
    using the same value of $\phi_{CHB}$. The dashed vertical line in
    (d) represents the value of $\phi_{TG}$ desired for accurate
    density matching, while the dashed vertical line in (a) provides
    the closeness of refractive index matching using the same value of
    $\phi_{TG}$.}
  \label{density-ri-match}
\end{figure}

The refractive index of CHB-DEC and TG-1P liquid mixtures was measured
using refractometer for varying fractions of CHB ($\phi_{CHB}$) and TG
($\phi_{TG}$), in their respective mixtures (see
figure~\ref{density-ri-match}a). The refractive index for CHB-DEC
mixture is nearly constant with varying $\phi_{CHB}$ while that for
TG-1P mixture shows a linear variation with $\phi_{TG}$. The
horizontal dashed line represents the reported value for the
refractive index of PMMA. In the vicinity of this value, $\phi_{CHB}$
and $\phi_{TG}$ were varied in small increments, respectively, in
CHB-DEC and TG-1P mixtures in order to determine the optimal binary
composition of each mixture at which the refractive index was closest
to PMMA. For each small variation, the suspension (of PMMA particles
in CHB-DEC or TG-1P liquid mixture) was placed in a transparent walled
rectangular cell and a thin laser line, of thickness about one-tenth
that of the particle diameter, was transmitted through one of the
faces of the cell, about $10$ particle diameters away from the other
orthogonal face of the cell. The laser line was imaged on a planar
surface on the other side of the cell placed orthogonal to the
incident laser line.  The values of $\phi_{CHB}$ and $\phi_{TG}$ in
their respective mixtures, which resulted in minimum scatter (measured
as the variation of intensity with the width of the scattered image)
yielded the optimal value of the liquid mixture composition (to the
order of third decimal place) for each constituent. This matching is
good enough to obtain near transparency upto about $20$ - $30$
particle diameters inside the sample.

The densities of the CHB-DEC and TG-1P liquid mixtures measured using
mass-volume method are shown in figure~\ref{density-ri-match}b for
varying $\phi_{CHB}$ and $\phi_{TG}$, in their respective
mixtures. The horizontal dashed line represents the reported value of
density for PMMA ($\approx 1.193$). In the vicinity of this value,
$\phi_{CHB}$ was varied (from $0.66$ to $0.71$) in very tiny
increments (inset (c) in figure~\ref{density-ri-match}b). As for the
refractive index matching procedure, for every small variation, the
suspension of PMMA particles in CHB-DEC liquid mixture, placed in a
transparent walled rectangular cell was illuminated using a thin laser
line.  Given the negligible variation of refractive index of CHB-DEC
mixture with its composition and its closeness to the refractive index
of particles, the sample is rendered nearly transparent. A small
amount of fluorescent dye (Carboxy-X-rhodamine succinimidyl ester
procured from Sigma-Aldrich) added to the CHB-DEC liquid mixture
fluoresces a plane in the sample showing (undyed) particles in that
plane as black shadows on a bright background. The negligible
variation of positions of different particles in any plane over a time
period of three hours provides the composition ($\phi_{CHB} = 0.686$)
which matches the density of CHB-DEC mixture with that of the particle
to the accuracy of three decimal places. Similar procedure (using Nile
Red from Sigma Aldrich as a fluorescent dye) was followed for TG-1P
mixture which yielded the composition ($\phi_{TG} = 0.865$) having the
density same as the particles to the accuracy of three decimal
places. Any further improvement in the accuracy is not possible due to
practical limitations in varying the concentration accurately
enough. 

It is to be noted that while the density matching composition for CHB-DEC
mixture is also close to the refractive index matching composition for
the same mixture (solid line in figure~\ref{density-ri-match}a), the
same does not hold for TG-1P mixture. The density matching composition
induces significant mis-match in the refractive indices between TG-1P
mixture and the particles (dashed line in
figure~\ref{density-ri-match}a). The primary importance during the gel
preparation is, however, given to accurate matching of densities of
liquid mixture and particles. Thus, suspensions of PMMA particles in
CHB-DEC mixture of composition $\phi_{CHB} = 0.686$, which is the
primary liquid, and containing tiny quantities of TG-1P mixture as the
secondary liquid of composition $\phi_{TG} = 0.865$ are exactly
density matched and are also transparent enough to be viewed to a
depth of about $20 - 30$ particle diameters inside the suspension. All
the measurements pertaining to density and refractive index matching
and laser imaging were performed in a room with temperature maintained
at $25 \pm 1$ deg C.

\bibliographystyle{spbasic}
\bibliography{ref-gel}

\end{document}